\documentclass[10pt,a4paper]{article}
\usepackage[T1]{fontenc}
\usepackage{amsmath,amsfonts, amssymb, amsbsy, mathrsfs, mathtools, bbm, enumitem}
\usepackage{titlesec}
\usepackage{float}
\usepackage{makeidx}
\usepackage{graphicx}
\usepackage{color}
\usepackage{url}
\usepackage{authblk}
\usepackage{multirow}
\usepackage{lscape}

\usepackage{mathpazo}
\usepackage[toc]{appendix}
\usepackage[hidelinks,colorlinks=true,linkcolor=red,citecolor=blue,urlcolor=magenta]{hyperref}
\usepackage[left=2.50cm, right=2.50cm, top=3.0cm, bottom=3.00cm]{geometry}

\usepackage{fancyhdr}

\newcommand{\email}[1]{\href{mailto:#1}{#1}}

\newcommand{\df}{\textrm{d}}
\newcommand{\Df}{\textrm{D}}

\newcommand{\e}{\textrm{e}}
\newcommand{\hf}{{\frac{1}{2}}}
\newcommand{\un}[1]{\underline{#1}}

\newcommand{\tr}{{^{{\rm (T)}}}}

\renewcommand{\bar}{\overline}

\titleformat{\paragraph}
{\normalfont\normalsize\bfseries}{\theparagraph}{1em}{}

\titleformat{\subparagraph}
{\normalfont\normalsize\bfseries}{\thesubparagraph}{1em}{}

\usepackage[square,numbers,sort&compress]{natbib}

\numberwithin{equation}{section}

\allowdisplaybreaks[1]

\begin{document}
	\setlength{\bibsep}{0pt}
	
	\title{ \textbf{Kinematics in Metric-Affine Geometry}}
	\author[]{Anish Agashe\thanks{\email{anagashe@smcm.edu}}}
	%\author[2]{Sai Madhav Modumudi\thanks{\email{saimadhav.modumudi@utdallas.edu}}}
	\affil[]{\small Department of Physics, St. Mary's College of Maryland,\\ 47645 College Dr, St. Mary's City, MD 20686 USA}
	%\affil[2]{\small Department of Physics, University of Texas at Dallas, 800 W Campbell Rd, Richardson, TX 75080}

	\maketitle
	
	\begin{abstract}
		In a given geometry, the kinematics of a congruence of curves is described by a set of three quantities called expansion, rotation, and shear. The equations governing the evolution of these quantities are referred to as kinematic equations. In this paper, the kinematics of congruence of curves in a metric-affine geometry are analysed. Without assuming an underlying theory of gravity, we derive a generalised form of the evolution equations for expansion, namely, Raychaudhuri equation (timelike congruences) and Sachs optical equation (null congruences). The evolution equations for rotation and shear of both timelike and null congruences are also derived. Generalising the deviation equation, we find that torsion and non-metricity contribute to a relative acceleration between neighbouring curves. We briefly discuss the interpretation of the expansion scalars and derive an equation governing angular diameter distances. The effects of torsion and non-metricity on the distances are found to be dependent on which curves are chosen as photon trajectories. We also show that the rotation of a hypersurface orthogonal congruence (timelike or null) is a purely non-Riemannian feature.
	\end{abstract}
	\section{Introduction}
	The general theory of relativity (GR) \cite{ein1,ein2} has been the basis of several models in astrophysics and cosmology. It has stood against the scrutiny of numerous experimental tests \cite{will}. Nevertheless, modifications and/or extensions to several of its underlying features have been proposed over the last century for various reasons.
	
	Given a geometry, two independent features can be endowed to it, namely, metric and connection. A metric formalises the notion of distances, while a connection formalises the notion of parallelism \cite{schout}. General relativity employs a geometrical framework in which that the latter is compatible with the former. That is, the connection can be expressed solely in terms of the metric and its (partial) derivatives. This is formally known as the metricity condition and such a connection is called metric compatible. In addition to being metric compatible, the connection in GR is also taken to be symmetric in two of its lower indices. Such connections are called torsion-free\footnote{The connections of GR that are both symmetric and metric compatible are commonly known as Levi-Civita connections or Christoffel symbols.}, torsion being the antisymmetric part of the connection. A geometry that is equipped with a torsion-free and metric compatible connection is termed as Riemannian\footnote{Throughout this paper, the signature is implicitly assumed to be Lorentzian. Therefore, by the term `Riemannian' what we really mean is (pseudo-)Riemannian.}. On the other hand, geometries in which any one of the aforementioned two conditions is violated are called non-Riemannian. The focus of our work here is a general non-Riemannian geometry, where both these conditions are violated at the same time.
	
	The interest in formulating a gravity theory using a non-Riemannian geometrical framework is not new. Within only a few years after general relativity was first proposed, Weyl considered geometries with non-metricity (but zero torsion)  in an attempt to unify electromagnetism with gravitation \cite{weyl1}. On the other hand, geometries with torsion (but zero non-metricity) were considered by Cartan \cite{cart0,cart1,cart2,cart3,cart4}. It has also been shown that there exist equivalent formalisms of GR in which gravitational effects can be attributed to either torsion or non-metricity, instead of the Riemann curvature tensor \cite{jim,capo4}. These are called teleparallel equivalent of general relativity (TEGR) \cite{haya,maluf,aldro,baha} and symmetric teleparallel equivalent of general relativity (STEGR) \cite{ferr,nest,capo1}, respectively. In general, given a geometry, any of the three geometrical objects, namely, the Riemann curvature, torsion, and non-metricity, (or a combination of them) can be used to encapsulate the gravitational effects. Depending on which of these three objects are considered, the landscape of non-Riemannian theories of gravity contain several special cases \cite{obu1}. The most general form of such theories, where all three objects are non-zero, is called metric-affine gravity (MAG) \cite{hehl1}. 
	
	With such non-Riemannian extensions of GR in mind, in this paper, we present the equations governing the mean kinematics of timelike and null curves in a general metric-affine geometry\footnote{By this, we mean the geometry corresponding to MAG. That is, a completely general non-Riemannian geometry with connection that is neither metric compatible nor torsion-free.}. Given a congruence of curves (timelike or null), its kinematics can be described in terms of the irreducible components of a tensor that governs the deviation between two infinitesimally separated curves. These components are known as expansion (the trace part), rotation (the antisymmetric part), and shear (the symmetric traceless part). Then, one can derive equations governing the evolution of these variables purely in terms of geometrical quantities\footnote{This means that no source of gravity (matter content) is assumed, nor a gravity theory (dynamics), making these equations kinematic in nature.}. The `Riemannian' counterpart of these equations has been extensively used in GR for topics like relativistic cosmology \cite{ellisbook}, black hole mechanics \cite{poisson}, gravitational lensing \cite{schneidbook}, space-time thermodynamics \cite{jacob}, singularity theorems \cite{hawkell}, gravitational collapse, and in numerous other physical contexts \cite{kar}. Therefore, it is important to have a generalised form of such kinematics which will enable one to investigate all these issues in the context of metric-affine geometries.
	
	To this end, we derive, to the best of our knowledge, the most general form of these kinematic equations. We work in $ n $-dimensions, do not assume a theory of gravity, and do not put any restrictions on the non-Riemannian variables (torsion and non-metricity). We also do not assume the geodesic condition, hence, making the results in this paper applicable to non-geodesic motion too. Recently, several generalisations of the kinematic equations in various non-Riemannian contexts have been presented. The effect of torsion on the kinematics was analysed in \cite{luz,hensh,dey1,dey2,dey3,capo2,wanas,cai,pasma,spez} while non-metricity (and torsion) was considered in \cite{lobo,iosth,ios,yang}. Here, we extend these previous works to include the equations for all three kinematic variables, including those for a congruence of null curves. We arrange our calculations such that all the kinematic variables and their corresponding equations are separated into a Riemannian part and a non-Riemannian part. This enables one to clearly identify the non-Riemannian contributions as `extra' terms, in addition to the usual Riemannian terms, and is helpful for finding a Riemannian limit for all the equations.
	
	We begin by describing the basic geometrical construction necessary to analyse kinematics, and underlining the differences between Riemannian and non-Riemannian cases. Working with this setup, we derive the deviation equation and find that both torsion and non-metricity contribute to a relative acceleration between infinitesimally separated curves. Then, we find the so called evolution tensor. The kinematic variables are defined as the irreducible parts of this tensor. Using these definitions, we proceed to find their evolution equations.
	
	After deriving the general form of the kinematic equations, we present a few special cases that are relevant to some particular theories of gravity. We also show that the expansion scalar does not retain its usual interpretation as the fractional rate of change of the volume (or area) of the congruence. Using the relationship between cosmological distances and the cross sectional area of null geodesic congruences, we derive an equation governing the angular diameter distances in metric-affine geometry. We find that distances will depend on which curves are chosen as photon trajectories. Considering particular types of torsion and non-metricity, we show that their effect on distances can be made indistinguishable by taking an ansatz relationship between the two. Further, we show that by demanding the congruence to be hypersurface orthogonal, one can determine the rotation tensor purely in terms of the non-Riemannian features of the geometry.
	
	The paper is arranged in the following manner: In section \ref{sec-einkin}, we present the standard derivation of kinematic equations in Riemannian geometry. We introduce some basics of metric-affine geometry in \ref{sec-mabasics}, and then, in section \ref{sec-makin}, we follow steps analogous to the Riemannian geometry to derive the generalised kinematic equations. After deriving these equations, we present some physically relevant special cases in section \ref{sec-spcase}, and derive an equation for angular diameter distance in section \ref{sec-cosdist}. Finally, in section \ref{sec-discuss}, we discuss the results in this paper and give some concluding remarks.
	
	The notation and convention used is as follows: objects associated with the Riemannian geometry are denoted by a `bar' over them. Greek indices run from $0$  to $n-1$ and Latin indices from $1$ to $n-1$. Indices with round brackets $ (~) $/square brackets $[~]$ are symmetrised/anti-symmetrised; and  underlined indices are not included in (anti-)symmetrisation. The partial derivative is denoted by $ \boldsymbol{\partial} $; covariant derivative is denoted by $\boldsymbol{\nabla}$; the directional derivative along a vector, $ \boldsymbol{X} $, is denoted by $ \boldsymbol{\Df_X} \equiv \boldsymbol{X\cdot\nabla} $; and the Lie derivative of a vector, $ \boldsymbol{Y} $, along another vector, $ \boldsymbol{X} $, denoted by $ \boldsymbol{\mathfrak{L}_X Y} \equiv \boldsymbol{X\cdot\partial Y - Y\cdot\partial X} $. The sign convention followed is the `Landau-Lifshitz Space-like Convention (LLSC)' \cite{mtw}. That is, the signature of the metric is taken to be Lorentzian $ (-,+,+,+,\dots) $, the Riemann curvature tensor is defined as, $ {R^\mu}_{\alpha\nu\beta}~=~2\partial_{[\nu}{\Gamma^\mu}_{\underline{\alpha}\beta]} + 2{\Gamma^\epsilon}_{\alpha[\beta}{\Gamma^\mu}_{\underline{\epsilon}\nu]} $ and $ R_{\mu\nu}~=~{R^\alpha}_{\mu\alpha\nu} $ is the Ricci tensor. The Ricci scalar is defined as $ R~=~{R^\mu}_\mu $.

	\section{Riemannian Geometry} \label{sec-einkin}
	The metric and the connection in a Riemannian geometry are not independent. The connection can be expressed in terms of the metric and its partial derivatives. Such a connection is called metric compatible. If the connection is also symmetric, it is referred to as the Levi-Civita connection. This is the geometry employed in GR and several of its (Riemannian) extensions \cite{clifrev,nojrev}.
	
	\subsection{Basic Definitions}
	Consider an arbitrary $ n $-dimensional Riemannian manifold with a metric, $ \boldsymbol{g} $, and Levi-Civita connection, $ \boldsymbol{\bar{\Gamma}} $. The Levi-Civita connection can be written in terms of metric as,
	\begin{equation}\label{levicivita}
		{\bar{\Gamma}^\rho}_{\alpha\beta} = \hf g^{\rho\sigma}\left(\partial_\alpha g_{\beta\sigma} + \partial_\beta g_{\sigma\alpha} - \partial_\sigma g_{\alpha\beta}\right)
	\end{equation}
	It is also symmetric in two of its indices,
	\begin{equation}\label{symmcon}
		{\bar{\Gamma}^\rho}_{\alpha\beta} = {\bar{\Gamma}^\rho}_{\beta\alpha} \quad {\rm or} \quad {\bar{\Gamma}^\rho}_{[\alpha\beta]} = 0
	\end{equation} 
	This is known as the torsion-free condition.
	
	The covariant derivative of an arbitrary rank tensor, $ \boldsymbol{V} $, is defined with respect to the Levi-Civita connection in the following manner,
	\begin{multline}\label{covder1}
		\bar{\nabla}_\rho {V^{\alpha_1 \alpha_2  \dots}}_{\beta_1 \beta_2 \dots} = \partial_\rho {V^{\alpha_1 \alpha_2  \dots}}_{\beta_1 \beta_2 \dots} + {\bar{\Gamma}^{\alpha_1}}_{\sigma\rho}{V^{\sigma \alpha_2  \dots}}_{\beta_1 \beta_2 \dots} + {\bar{\Gamma}^{\alpha_2}}_{\sigma\rho}{V^{ \alpha_1 \sigma  \dots}}_{\beta_1 \beta_2 \dots} + \dots \\- {\bar{\Gamma}^\sigma}_{\beta_1\rho}{V^{ \alpha_1 \alpha_2  \dots}}_{\sigma \beta_2 \dots} - {\bar{\Gamma}^\sigma}_{\beta_2\rho}{V^{ \alpha_1 \alpha_2 \dots}}_{\beta_1 \sigma \dots} - \dots
	\end{multline}
	{With the definition of covariant derivative, one can show that,}
	\begin{equation}\label{metricity2}
		\bar{\nabla}_\alpha g_{\beta\sigma} = 0
	\end{equation}
	This is known as the metricity condition. {Conversely, equation \eqref{levicivita} can be derived by using the metricity and torsion-free conditions together.} Later in this paper, we will see the consequences of violation of both these conditions.
	
	Furthermore, the Ricci identity is given by,
	\begin{equation}\label{ricciiden}
		\left(\overline{\nabla}_\alpha\bar{\nabla}_\beta - \bar{\nabla}_\beta\bar{\nabla}_\alpha \right)V^\rho = {\bar{R}^\rho}_{\sigma\alpha\beta}V^\sigma
	\end{equation}
	where, $ {\bar{R}^\rho}_{\sigma\alpha\beta} $, is the Riemann curvature tensor defined in terms of the Levi-Civita connection as,
	\begin{equation}\label{riemcurv}
		{\bar{R}^\rho}_{\sigma\alpha\beta} = \partial_\alpha {\bar{\Gamma}^\rho}_{\sigma\beta} - \partial_\beta {\bar{\Gamma}^\rho}_{\sigma\alpha} + {\bar{\Gamma}^\rho}_{\mu\alpha}{\bar{\Gamma}^\mu}_{\sigma\beta} - {\bar{\Gamma}^\rho}_{\mu\beta}{\bar{\Gamma}^\mu}_{\sigma\alpha}
	\end{equation}
	The Ricci tensor is defined as the trace of the Riemann curvature tensor,
	\begin{equation}\label{ricciten}
		\bar{R}_{\sigma\beta} = g^{\rho\alpha}\bar{R}_{\rho\sigma\alpha\beta} =  {\bar{R}^\alpha}_{\sigma\alpha\beta}
	\end{equation}
	And the Ricci scalar is defined as,
	\begin{equation}\label{ricciscalar}
		\bar{R} = g^{\sigma\beta}\bar{R}_{\sigma\beta} = {\bar{R}^{\alpha\beta}}_{\alpha\beta}
	\end{equation}
	
	\subsection{Kinematics in Riemannian Geometry}
	To analyse kinematics of curves in Riemannian geometry, we start by considering a congruence of curves. Given an open region, $ \mathcal{O} $, in the manifold, a congruence in $ \mathcal{O} $ is a family of curves such that through each point in $ \mathcal{O} $, there passes one and only one curve from this family \cite{poisson}. Let these curves be parametrised by a parameter, $ t $. Then, the vector field, $ u^\alpha = \frac{\partial x^\alpha}{\partial t} $, is tangent to the curves, and we have,
	\begin{equation}\label{vecnorm}
		u^\alpha u_\alpha = \epsilon
	\end{equation}
	where, for $ \epsilon = -1, 0, +1 $, the curves are timelike, null, and spacelike, respectively.
	
	To determine how the congruence evolves, we consider another family of curves along each of which, the parameter, $ t $, remains fixed. These curves can be considered to be parametrised by another parameter, say, $ s $. Then, the vector field, $ \xi^\alpha = \frac{\partial x^\alpha}{\partial s} $, is now the tangent to these `$ t $-constant' curves. It is easy to see that the Lie derivatives of the two tangent vectors with respect to each other vanish,
	\begin{equation}\label{lieder1}
		\mathfrak{L}_u \xi = 0 = \mathfrak{L}_\xi u
	\end{equation}
	From here, it follows that,
	\begin{equation}\label{deviation}
		\bar{\Df}_u \xi^\alpha = \bar{\Df}_\xi u^\alpha  \Leftrightarrow u^\beta\bar{\nabla}_\beta \xi^\alpha = \xi^\beta\bar{\nabla}_\beta u^\alpha 
	\end{equation}
	
	Since the $ t $-constant curves span `across' the congruence, the evolution of the tangent vector, $ \xi^\alpha $, tells us about the deviation of two neighbouring curves in the congruence under investigation. Hence, $ \xi^\alpha $ is termed as `deviation vector'. To add weight to this interpretation, it can be shown that, 
	\begin{equation}
		\bar{\Df}_u\left(\xi^\alpha u_\alpha \right) = \xi^\alpha \bar{a}_\alpha 
		\Rightarrow \frac{\df }{\df t} \left(\xi^\alpha u_\alpha \right) = \xi^\alpha \bar{a}_\alpha \label{devdirecderi}
	\end{equation} 
	where, the acceleration, $ \boldsymbol{\bar{a}} $, is defined as,
	\begin{equation}\label{accein}
		\bar{a}^\alpha = \bar{\Df}_u u^\alpha
	\end{equation}
	Note that, $\bar{a}_\alpha =  g_{\alpha\beta} \bar{a}^\beta = \bar{\Df}_u u_\alpha $. In the case of Riemannian geometry, assuming affine parametrisation, a curve being a geodesic corresponds to zero acceleration, $ \bar{a}^\alpha = 0 $. Therefore, for geodesics, the right hand side of equation \eqref{devdirecderi} becomes zero. Then, the parametrisation of $ \boldsymbol{u} $ and $ \boldsymbol{\xi} $ can always be chosen such that their product is zero (instead of a constant). Therefore, the two tangent vectors are orthogonal for geodesic congruences. This means that the deviation vector captures the transverse properties of the congruence.
	
	\subsubsection{Deviation Equation} 
	Using equation \eqref{deviation}, one can derive the so-called `deviation equation' for a congruence of curves,
	\begin{equation}\label{deviation3}
		\bar{\Df}^2_u \xi^\alpha = {\bar{R}^\alpha}_{\beta\rho\sigma}u^\beta u^\rho \xi^\sigma + \xi^\sigma\bar{\nabla}_\sigma\bar{a}^\alpha
	\end{equation}
	For geodesic motion ($\boldsymbol{\bar{a}} = 0$), the above equation is known as the geodesic deviation equation. Using this equation, one can see that Riemann curvature leads to a relative acceleration between two initially parallel geodesics. Physically, this relative acceleration can be interpreted to be due to the tidal effects from gravitational force.
	\subsubsection{Kinematics of Timelike Curves}
	In this section, we derive the irreducible kinematics of a congruence of timelike curves ($ u^\alpha u_\alpha = -1 $). Let us rewrite equation \eqref{deviation} as,
	\begin{equation}\label{evolein1}
		\bar{\Df}_u\xi^\alpha = {\bar{B}^\alpha}_\beta\xi^\beta
	\end{equation}
	where, 
	\begin{equation}\label{evolein2}
		{\bar{B}^\alpha}_\beta = \bar{\nabla}_\beta u^\alpha
	\end{equation}
	We will term $ \boldsymbol{\bar{B}} $ as the `evolution tensor' since it governs the evolution of the deviation vector\footnote{In Riemannian geometry, this evolution tensor is simply the gradient of the tangent vector (4-velocity gradient when $ n = 4 $). Therefore, one can write the kinematics directly from the Ricci identity. However, we will see that it is not the case when we go to metric-affine geometry.}. It is the fractional rate of change of the deviation vector. Therefore, the evolution equations for the irreducible parts of this tensor provide us with the kinematics of the congruence. However, we are interested in the transverse properties of the  congruence and for that we need the component of the evolution tensor that is transverse to the congruence. To find this, we define a transverse metric as,
	\begin{equation}\label{transmetricein}
		h_{\alpha\beta} = g_{\alpha\beta} + u_\alpha u_\beta
	\end{equation}
	where, using $ u^\alpha u_\alpha = -1 $, it is easy to check that, $ u^\alpha h_{\alpha\beta}~=~0~=~h_{\alpha\beta}u^\beta $. Also, it follows that, $ {h^\alpha}_\sigma {h^\sigma}_\beta = {h^\alpha}_\beta $ and $ h^{\alpha\beta}h_{\alpha\beta} = {h^\alpha}_\alpha = n-1 $.
	
	Using this, the evolution tensor can be divided in a transverse part and a longitudinal part,
	\begin{align}
		^{{\rm (T)}}\bar{B}_{\alpha\beta} &= {h^\rho}_\alpha {h^\sigma}_\beta \bar{B}_{\rho\sigma} = \bar{B}_{\alpha\beta} + \bar{a}_\alpha u_\beta \label{transevolein}\\
		^{{\rm (L)}}\bar{B}_{\alpha\beta} &= \bar{B}_{\alpha\beta} - ^{{\rm (T)}}\bar{B}_{\alpha\beta} \label{evollong}
	\end{align}
	where, it is clear that, $ u^\alpha {^{{\rm (T)}}\bar{B}_{\alpha\beta}} = 0 = ^{{\rm (T)}}\bar{B}_{\alpha\beta} u^\beta $.
	
	Then, the kinematic quantities are defined as the trace, anti-symmetric, and symmetric traceless parts of $ ^{{\rm (T)}}\bar{B}_{\alpha\beta} $,
	\begin{align}
		\bar{\theta} &= g^{\alpha\beta}{^{{\rm (T)}}\bar{B}_{\alpha\beta}} \label{expeintim}\\
		\bar{\omega}_{\alpha\beta} &= ^{{\rm (T)}}\bar{B}_{[\alpha\beta]} \label{roteintim}\\
		\bar{\sigma}_{\alpha\beta} &= ^{{\rm (T)}}\bar{B}_{(\alpha\beta)} - \frac{1}{{h^\rho}_\rho}\bar{\theta} h_{\alpha\beta} \label{sheareintim}
	\end{align}
	where, $\bar{\theta}$ is the expansion, $ \boldsymbol{\bar{\omega}} $ the is rotation, and $ \boldsymbol{\bar{\sigma}} $ is the shear of the congruence. Using these definitions and equation \eqref{transevolein}, we can write,
	\begin{equation}\label{expein}
		\bar{\theta} = {\bar{B}^\alpha}_\alpha + \bar{a}^\alpha u_\alpha
	\end{equation}
	\begin{equation}\label{rotein}
		\bar{\omega}_{\alpha\beta} = \bar{B}_{[\alpha\beta]} + \bar{a}_{[\alpha}u_{\beta]}
	\end{equation}
	\begin{equation}\label{shearein}
		\bar{\sigma}_{\alpha\beta} = \bar{B}_{(\alpha\beta)} + \bar{a}_{(\alpha}u_{\beta)} - \frac{1}{n-1} \bar{\theta}h_{\alpha\beta}
	\end{equation}
	The evolution tensor $ {^{{\rm (T)}}\bar{B}_{\alpha\beta}} $ is determined completely by these three quantities,
	\begin{equation}\label{evolcomp}
		{^{{\rm (T)}}\bar{B}_{\alpha\beta}} = {^{{\rm (T)}}\bar{B}_{(\alpha\beta)}} + {^{{\rm (T)}}\bar{B}_{[\alpha\beta]}} = \frac{1}{n-1}\bar{\theta} h_{\alpha\beta} + \bar{\omega}_{\alpha\beta} + \bar{\sigma}_{\alpha\beta}
	\end{equation}
	It is useful to define scalar quantities, $ \bar{\omega}^2 = \bar{\omega}^{\alpha\beta}\bar{\omega}_{\alpha\beta} $ and $ \bar{\sigma}^2 = \bar{\sigma}^{\alpha\beta}\bar{\sigma}_{\alpha\beta}$, which gives us,
	\begin{equation}\label{evolsquare}
		{^{{\rm (T)}}}\bar{B}^{\alpha\beta}\tr\bar{B}_{\alpha\beta} = \frac{1}{n-1}\bar{\theta}^2 + \bar{\omega}^2 + \bar{\sigma}^2 
	\end{equation}
	
	Using these definitions, we can write the kinematic equations for a congruence of timelike curves,
	\begin{equation}\label{einkin1}
		\bar{\Df}_u\bar{\theta} = -\frac{1}{n-1} \bar{\theta}^2 + \bar{\omega}^2 - \bar{\sigma}^2 - \bar{R}_{\alpha\beta}u^\alpha u^\beta \\+ \bar{\Df}_u\left(\bar{a}^\alpha u_\alpha\right) + \bar{\nabla}_\alpha\bar{a}^\alpha
	\end{equation}
	\begin{equation}\label{einkin2}
		\bar{\Df}_u\bar{\omega}_{\alpha\beta} = -\frac{2}{n-1}\bar{\theta}\bar{\omega}_{\alpha\beta} - \bar{\sigma}_{\alpha\rho}{\bar{\omega}^\rho}_\beta - \bar{\omega}_{\alpha\rho}{\bar{\sigma}^\rho}_\beta\\ - u_{[\alpha} \bar{\Df}_u\bar{a}_{\beta]} - u_{[\alpha}\bar{\nabla}_{\un{\rho}}u_{\beta]}\bar{a}^\rho - \bar{\nabla}_{[\alpha} \bar{a}_{\beta]} 
	\end{equation}
	\begin{multline}\label{einkin3}
		\bar{\Df}_u\bar{\sigma}_{\alpha\beta} = - \frac{2}{n-1}\bar{\theta}\bar{\sigma}_{\alpha\beta} -\bar{\omega}_{\alpha\rho}{\bar{\omega}^\rho }_\beta - \bar{\sigma}_{\alpha\rho}{\bar{\sigma}^\rho }_\beta + \frac{1}{n-1}h_{\alpha\beta}\left( \bar{\sigma}^2 - \bar{\omega}^2 \right) \\+ \frac{1}{n-1}h_{\alpha\beta}\bar{R}_{\rho\sigma}u^\rho u^\sigma\ - \bar{R}_{\alpha\rho\beta\sigma}u^\rho u^\sigma - \frac{2}{n-1}\bar{\theta}u_{(\alpha}\bar{a}_{\beta)} - \frac{1}{n-1}h_{\alpha\beta}\left\{\bar{\Df}_u(\bar{a}^\alpha u_\alpha) + \bar{\nabla}_\rho\bar{a}^\rho\right\} \\ + u_{(\alpha}\bar{\nabla}_{\un{\rho}}u_{\beta)}\bar{a}^\rho + u_{(\alpha}\bar{\Df}_u\bar{a}_{\beta)} + \bar{\nabla}_{(\alpha} \bar{a}_{\beta)} + \bar{a}_\alpha\bar{a}_\beta 
	\end{multline}
	The equation for the expansion above is known as the Raychaudhuri equation \cite{ray1,poisson}. The equations above are valid for a congruence of arbitrary timelike curves (no geodesic condition has been assumed). The kinematics of a congruence of non-affinely parametrised timelike geodesics can be found by substituting $ \bar{a}^\alpha = \kappa u^\alpha $, and that of affinely parametrised geodesics by taking $ \bar{a}^\alpha  = 0 $.
	
	\subsubsection{Kinematics of Null Curves}
	For deriving the kinematics of a congruence of null curves, we again start with equations \eqref{evolein1} and \eqref{evolein2}. We will relabel the tangent vector field for null curves as, $ k^\alpha k_\alpha = 0 $, such that the evolution tensor becomes, $ {\bar{B}^\alpha}_\beta = \bar{\nabla}_\beta k^\alpha $. It is clear that the evolution tensor is not transverse to the congruence and hence we need to separate out its transverse part. For this, we need the transverse metric which in this case is defined as,
	\begin{equation}\label{nulltransmetric}
		h_{\alpha\beta} = g_{\alpha\beta} + k_\alpha X_\beta + X_\alpha k_\beta
	\end{equation}
	where, $ \boldsymbol{X} $ is an auxiliary null vector field, $ X^\alpha X_\alpha = 0 $, and is normalised such that, $ X^\alpha k_\alpha = -1 $. Using the above definition of the transverse metric, it can be readily shown that, $ k^\alpha h_{\alpha\beta} = 0 = h_{\alpha\beta}k^\beta $. Also, it follows that, $ {h^\alpha}_\sigma {h^\sigma}_\beta = {h^\alpha}_\beta $ and $ h^{\alpha\beta}h_{\alpha\beta} = {h^\alpha}_\alpha = n-2 $. Then, the transverse part of the evolution tensor is defined in same manner as equation \eqref{transevolein},
	\begin{equation}\label{transevolnull}
		\tr\bar{B}_{\alpha\beta} = {h^\rho}_\alpha {h^\sigma}_\beta \bar{B}_{\rho\sigma} = \bar{B}_{\alpha\beta} + \bar{a}_\alpha X_\beta + \bar{B}_{\alpha\sigma}X^\sigma k_\beta + k_\alpha\bar{B}_{\rho\beta}X^\rho + k_\alpha X_\beta X^\rho \bar{a}_\rho + k_\alpha k_\beta \bar{B}_{\rho\sigma}X^\rho X^\sigma
	\end{equation}
	The definitions of the three kinematic variables (equations \eqref{expeintim} - \eqref{sheareintim}) follow from here and we relabel them as, $ \bar{\Theta} $, $ \bar{\Omega} $, and $ \bar{\Sigma} $ to avoid confusion with their timelike counterparts. The kinematic variables are given by,
	\begin{equation}
		\bar{\Theta} = {\bar{B}^\alpha}_\alpha + X^\alpha \bar{a}_\alpha
	\end{equation}
	\begin{equation}
		\bar{\Omega}_{\alpha\beta} = \bar{B}_{[\alpha\beta]} + \bar{B}_{[\alpha\sigma]}k_\beta X^\sigma - \bar{B}_{[\beta\sigma]}k_\alpha X^\sigma + \bar{a}_{[\alpha}X_{\beta]} + k_{[\alpha}X_{\beta]}\bar{a}_\rho X^\rho
	\end{equation}
	\begin{multline}
		\bar{\Sigma}_{\alpha\beta} = \bar{B}_{(\alpha\beta)} + \bar{B}_{(\alpha\sigma)}k_\beta X^\sigma + \bar{B}_{(\beta\sigma)}k_\alpha X^\sigma + \bar{a}_{(\alpha}X_{\beta)} + k_\alpha k_\beta \bar{B}_{\rho\sigma}X^\rho X^\sigma  \\+ k_{(\alpha}X_{\beta)}\bar{a}_\rho X^\rho - \frac{1}{n-2} \bar{\Theta}h_{\alpha\beta}
	\end{multline}
	
	Then, the kinematic equations will be given by,
	\begin{multline}\label{einkinnull1}
		\bar{\Df}_k \bar{\Theta} = -\frac{1}{n-2} \bar{\Theta}^2 + \bar{\Omega}^2 - \bar{\Sigma}^2 - \bar{R}_{\alpha\beta}k^\alpha k^\beta + \bar{\Df}_k \left(\bar{a}^\alpha X_\alpha\right) + 2\bar{a}^\beta \bar{\nabla}_\beta k^\alpha X_\alpha \\ + \bar{\nabla}_\alpha \bar{a}^\alpha - \bar{a}^\alpha X_\alpha \bar{a}^\beta X_\beta
	\end{multline}
	The equation for the rotation can be written in the following form,
	\begin{multline}\label{einkinnull2}
		\bar{\Df}_k \bar{\Omega}_{\alpha\beta} =  \bar{\Df}_k \bar{B}_{[\alpha\beta]} +  \bar{\Df}_k \bar{B}_{[\alpha\rho]}k_\beta X^\rho - k_\alpha \bar{\Df}_k\bar{B}_{[\beta\rho]}X^\rho + \bar{B}_{[\alpha\rho]}k_\beta\bar{\Df}_k X^\rho - k_\alpha\bar{B}_{[\beta\rho]}\bar{\Df}_k X^\rho \\+ \bar{\Df}_k\left(\bar{a}_{[\alpha}X_{\beta]}\right) + \bar{\Df}_k \left( k_{[\alpha}X_{\beta]} X^\rho\bar{a}_\rho\right) + \bar{B}_{[\alpha\rho]}\bar{a}_\beta X^\rho - \bar{a}_\alpha \bar{B}_{[\beta\rho]}X^\rho
	\end{multline}
	where, we have,
	\begin{multline}
		\bar{\Df}_k\bar{B}_{[\alpha\beta]} = -\frac{2}{n-2}\bar{\Theta}\ \bar{\Omega}_{\alpha\beta} - \bar{\Omega}_{\alpha\rho}{\bar{\Sigma}^\rho}_\beta - \bar{\Sigma}_{\alpha\rho}{\bar{\Omega}^\rho}_\beta + \bar{B}_{[\alpha\un{\rho}}k_{\beta]}{\bar{B}^\rho}_\sigma X^\sigma + k_{[\alpha}\bar{B}_{\un{\rho}\beta]}{\bar{B}_\sigma}^\rho X^\sigma \\ + \bar{B}_{[\alpha\un{\rho}}X_{\beta]}\bar{a}^\rho - \bar{\nabla}_{[\alpha}\bar{a}_{\beta]} + \left( \bar{B}_{[\alpha\un{\rho}}k_{\beta]} + k_{[\alpha}\bar{B}_{\un{\rho}\beta]} \right)X^\rho \bar{a}_\sigma X^\sigma + \left( \bar{B}_{[\alpha\un{\rho}}\bar{a}_{\beta]} + \bar{a}_{[\alpha}\bar{B}_{\un{\rho}\beta]} \right)X^\rho \\ + k_{[\alpha}X_{\beta]} \left( \bar{B}_{\rho\sigma}X^\rho\bar{a}^\sigma + X_\rho X_\sigma \bar{a}^\rho\bar{a}^\sigma \right) + \bar{a}_{[\alpha}X_{\beta]}X_\rho\bar{a}^\rho 
	\end{multline}
	and,
	\begin{equation}\label{antisymevol}
		\bar{B}_{[\alpha\beta]} = \bar{\Omega}_{\alpha\beta} - \left( \bar{B}_{[\alpha\un{\rho}}k_{\beta]} + k_{[\alpha}\bar{B}_{\un{\rho}\beta]} \right)X^\rho - \bar{a}_{[\alpha}X_{\beta]} - k_{[\alpha}X_{\beta]}X^\rho\bar{a}_\rho
	\end{equation}
	Similarly, the equation for shear can be written as,
	\begin{multline}\label{einkinnull3}
		\bar{\Df}_k \bar{\Sigma}_{\alpha\beta} = \bar{\Df}_k\bar{B}_{(\alpha\beta)} - \frac{1}{n-2}h_{\alpha\beta}\bar{\Df}_k\bar{\Theta} + \bar{\Df}_k \left[ \left\{ \bar{B}_{(\alpha\rho)}k_\beta + k_\alpha\bar{B}_{(\beta\rho)} \right\}X^\rho \right] \\+ k_\alpha k_\beta \bar{\Df}_k\left( \bar{B}_{(\rho\sigma)}X^\rho X^\sigma \right) + \bar{\Df}_k\left(\bar{a}_{(\alpha}X_{\beta)} + k_{(\alpha}X_{\beta)}X^\rho\bar{a}_\rho\right) \\+ 2\bar{a}_{(\alpha}k_{\beta)}\left( \bar{B}_{(\rho\sigma)}X^\rho X^\sigma - \frac{2}{n-2}\bar{\Theta} \right)
	\end{multline}
	where, we have,
	\begin{multline}
		\bar{\Df}_k\bar{B}_{(\alpha\beta)} - 	\frac{1}{n-2}h_{\alpha\beta}\bar{\Df}_k\bar{\Theta} = -\frac{2}{n-2}\bar{\Theta}\ \bar{\Sigma}_{\alpha\beta} - \bar{\Omega}_{\alpha\rho}{\bar{\Omega}^\rho}_\beta - \bar{\Sigma}_{\alpha\rho}{\bar{\Sigma}^\rho}_\beta - \bar{R}_{\alpha\rho\beta\sigma}k^\rho k^\sigma \\ + \bar{B}_{(\alpha\un{\rho}}k_{\beta)}{\bar{B}^\rho}_\sigma X^\sigma + k_\alpha k_\beta \bar{B}_{\rho\sigma}{\bar{B}^\rho}_\epsilon X^\sigma X^\epsilon + 2\bar{a}_{(\alpha}k_{\beta)}\bar{B}_{\rho\sigma}X^\rho X^\sigma  \\ + \bar{B}_{(\alpha\un{\rho}}X_{\beta)}\bar{a}^\rho + \left( \bar{B}_{(\alpha\un{\rho}}k_{\beta)} + k_{(\alpha}\bar{B}_{\un{\rho}\beta)} \right)X^\rho X^\sigma \bar{a}_\sigma + \left( \bar{a}_{(\alpha}\bar{B}_{\un{\rho}\beta)} + \bar{B}_{(\alpha\un{\rho}}\bar{a}_{\beta)} \right)X^\rho \\  + k_{(\alpha}\bar{B}_{\un{\rho}\beta)}{\bar{B}_\sigma}^\rho X^\sigma + k_{(\alpha}X_{\beta)}\left( \bar{B}_{\rho\sigma} + \bar{a}_\rho X_\sigma \right) X^\rho\bar{a}_\sigma + 2k_\alpha k_\beta \bar{B}_{\rho\sigma}X^\rho X^\sigma X^\epsilon\bar{a}_\epsilon + \bar{\nabla}_{(\alpha}\bar{a}_{\beta)} \\ - \frac{1}{n-2}h_{\alpha\beta}\left\{ \bar{\Omega}^2 - \bar{\Sigma}^2 - \bar{R}_{\alpha\beta}k^\alpha k^\beta + \bar{\Df}_k \left(\bar{a}^\alpha X_\alpha\right) + 2\bar{a}^\beta \bar{\nabla}_\beta k^\alpha X_\alpha + \bar{\nabla}_\alpha \bar{a}^\alpha - \bar{a}^\alpha X_\alpha \bar{a}^\beta X_\beta \right\} 
	\end{multline}
	The null counterpart of the Raychaudhuri equation was first given by Sachs \cite{sachs1,sachs2}. For this reason, it is sometimes called Sachs (optical) equation. The equations above are valid for a congruence of arbitrary null curves (no geodesic condition has been assumed). The kinematics of a congruence of non-affinely parametrised null geodesics can be found by substituting $ \bar{a}^\alpha = \kappa k^\alpha $, and that of affinely parametrised geodesics by taking $ \bar{a}^\alpha  = 0 $.
	
	\section{Metric-Affine Geometry} \label{sec-mabasics}
	Unlike the Riemannian geometry, the metric and the connection in a metric-affine geometry are independent objects. {That is, both the metricity and torsion-free conditions (equations \eqref{metricity2} and \eqref{symmcon}) do not hold and hence one cannot express the connection in terms of metric as done in equation \eqref{levicivita}.} Therefore, the non-Riemannian nature of the geometry is essentially determined by the connection following or not following certain conditions.
	
	\subsection{Basic Definitions}
	Consider an arbitrary $ n $-dimensional metric-affine manifold with a metric, $ \boldsymbol{g} $, and a connection, $ \boldsymbol{\Gamma} $. Then, torsion, $ \boldsymbol{T} $, is defined as the antisymmetric part of the connection,
	\begin{equation}\label{torsion}
		{T^\rho}_{\alpha\beta} = {\Gamma^\rho}_{[\alpha\beta]} = \hf \left( {\Gamma^\rho}_{\alpha\beta} - {\Gamma^\rho}_{\beta\alpha}\right)
	\end{equation}
	
	The covariant derivative is now defined with respect to this connection,
	\begin{multline}\label{covder2}
		{\nabla}_\rho {V^{\alpha_1 \alpha_2  \dots}}_{\beta_1 \beta_2 \dots} = \partial_\rho {V^{\alpha_1 \alpha_2  \dots}}_{\beta_1 \beta_2 \dots} + {{\Gamma}^{\alpha_1}}_{\sigma\rho}{V^{\sigma \alpha_2  \dots}}_{\beta_1 \beta_2 \dots} + {{\Gamma}^{\alpha_2}}_{\sigma\rho}{V^{ \alpha_1 \sigma  \dots}}_{\beta_1 \beta_2 \dots} + \dots \\- {{\Gamma}^\sigma}_{\beta_1\rho}{V^{ \alpha_1 \alpha_2  \dots}}_{\sigma \beta_2 \dots} - {{\Gamma}^\sigma}_{\beta_2\rho}{V^{ \alpha_1 \alpha_2  \dots}}_{\beta_1 \sigma \dots} - \dots
	\end{multline}
	Then, non-metricity, $ \boldsymbol{Q} $, is defined as,
	\begin{equation}\label{nonmetricity}
		Q_{\alpha\beta\sigma} = \nabla_\alpha g_{\beta\sigma} = \partial_\alpha g_{\beta\sigma} - {\Gamma^\rho}_{\beta\alpha} g_{\rho\sigma} - {\Gamma^\rho}_{\sigma\alpha} g_{\beta\rho} = \partial_\alpha g_{\beta\sigma} - \Gamma_{\sigma\beta\alpha} - \Gamma_{\beta\sigma\alpha}
	\end{equation}
	It is easy to see that non-metricity is symmetric in two of its indices,
	\begin{equation}\label{nonmetsymm}
		Q_{\alpha(\beta\sigma)} = \hf \left( Q_{\alpha\beta\sigma} + Q_{\alpha\sigma\beta} \right) = Q_{\alpha\beta\sigma}
	\end{equation}
	
	The connection can be decomposed in the following manner,
	\begin{equation}\label{distor1}
		{\Gamma^{\rho}}_{\alpha\beta} = {\bar{\Gamma}^\rho}_{\alpha\beta} + {N^{\rho}}_{\alpha\beta} 
	\end{equation}
	where, $ \boldsymbol{\bar{\Gamma}} $ is the Riemannian part (Levi-Civita connection), and $ \boldsymbol{N} $ is the so-called distortion tensor that encapsulates the deviation from the Riemannian nature of geometry. The distortion is given in terms of the torsion and non-metricity as,
	\begin{equation}\label{distor2}
		{N^{\rho}}_{\alpha\beta} = \left( {T^{\rho}}_{\alpha\beta} - {T_\beta}^\rho\! _\alpha + {T_{\alpha\beta}}^\rho \right) + \hf \left( {Q^{\rho}}_{\alpha\beta} - {Q_\beta}^\rho\! _\alpha - {Q_{\alpha\beta}}^\rho \right)
	\end{equation}
	The combination of the torsion above is called the contorsion torsion, $ \boldsymbol{K} $, given by,
	\begin{equation}\label{contor}
		{K^{\rho}}_{\alpha\beta} = {T^{\rho}}_{\alpha\beta} - {T_\beta}^\rho\! _\alpha + {T_{\alpha\beta}}^\rho
	\end{equation}
	The combination of the non-metricity is sometimes referred to as deformation tensor, $ \boldsymbol{P} $, given by,
	\begin{equation}\label{qcomb}
		{P^{\rho}}_{\alpha\beta} = \hf \left( {Q^{\rho}}_{\alpha\beta} - {Q_\beta}^\rho\! _\alpha - {Q_{\alpha\beta}}^\rho \right)
	\end{equation}
	Then, using these two definitions, the distortion tensor can be written as,
	\begin{equation}\label{distor3}
		{N^{\rho}}_{\alpha\beta} = {K^{\rho}}_{\alpha\beta} + {P^{\rho}}_{\alpha\beta}
	\end{equation}
	Using equations \eqref{nonmetricity} and \eqref{distor1}, we can write,
	\begin{equation}\label{distornonmet}
		N_{(\beta\sigma)\alpha} = -\hf Q_{\alpha\beta\sigma}
	\end{equation}
	Moreover, it is easy to check that,
	\begin{equation}\label{distortor}
		N_{\rho[\alpha\beta]} = T_{\rho\alpha\beta}
	\end{equation}
	Using the distortion tensor, the covariant derivative can be divided into Riemannian and non-Riemannian parts,
	\begin{multline}\label{covder3}
		{\nabla}_\rho {V^{\alpha_1 \alpha_2  \dots}}_{\beta_1 \beta_2 \dots} = \bar{\nabla}_\rho {V^{\alpha_1 \alpha_2  \dots}}_{\beta_1 \beta_2 \dots} + {{N}^{\alpha_1}}_{\sigma\rho}{V^{\sigma \alpha_2  \dots}}_{\beta_1 \beta_2 \dots} + {{N}^{\alpha_2}}_{\sigma\rho}{V^{ \alpha_1 \sigma  \dots}}_{\beta_1 \beta_2 \dots} + \dots \\- {{N}^\sigma}_{\beta_1\rho}{V^{ \alpha_1 \alpha_2  \dots}}_{\sigma \beta_2 \dots}  - {{N}^\sigma}_{\beta_2\rho}{V^{ \alpha_1 \alpha_2  \dots}}_{\beta_1 \sigma \dots} - \dots
	\end{multline}
	
	Further, the Ricci identity now becomes,
	\begin{equation}\label{ricciidenma}
		\left({\nabla}_\alpha{\nabla}_\beta - {\nabla}_\beta{\nabla}_\alpha \right)V^\rho = {{R}^\rho}_{\sigma\alpha\beta}V^\sigma + 2 {T^\sigma}_{\alpha\beta} \nabla_\sigma V^\rho 
	\end{equation}
	where, $ {{R}^\rho}_{\sigma\alpha\beta} $, is the Riemann curvature tensor defined in terms of the connection of the metric-affine geometry,
	\begin{equation}\label{riemcurvma}
		{{R}^\rho}_{\sigma\alpha\beta} = \partial_\alpha {{\Gamma}^\rho}_{\sigma\beta} - \partial_\beta {{\Gamma}^\rho}_{\sigma\alpha} + {{\Gamma}^\rho}_{\mu\alpha}{{\Gamma}^\mu}_{\sigma\beta} - {{\Gamma}^\rho}_{\mu\beta}{{\Gamma}^\mu}_{\sigma\alpha}
	\end{equation}
	Using equation \eqref{distor1}, the curvature tensor can be decomposed into Riemannian and non-Riemannian parts,
	\begin{equation}\label{curvdecomp}
		{{R}^\rho}_{\sigma\alpha\beta} = {\bar{R}^\rho}_{\sigma\alpha\beta} + {{W}^\rho}_{\sigma\alpha\beta} + {{S}^\rho}_{\sigma\alpha\beta}
	\end{equation}
	where, the tensors, $ \boldsymbol{W} $ and $ \boldsymbol{S} $, are defined as,
	\begin{equation}
		{{W}^\rho}_{\sigma\alpha\beta} = \partial_\alpha {{N}^\rho}_{\sigma\beta} - \partial_\beta {{N}^\rho}_{\sigma\alpha} + {{N}^\rho}_{\mu\alpha}{{N}^\mu}_{\sigma\beta} - {{N}^\rho}_{\mu\beta}{{N}^\mu}_{\sigma\alpha} \label{mdef}
	\end{equation}
	\begin{equation}
		{{S}^\rho}_{\sigma\alpha\beta} = \left({\bar{\Gamma}^\rho}_{\mu\alpha}{{N}^\mu}_{\sigma\beta} - {{N}^\rho}_{\mu\beta}{\bar{\Gamma}^\mu}_{\sigma\alpha}\right) + \left({N^\rho}_{\mu\alpha}{{\bar{\Gamma}}^\mu}_{\sigma\beta} - {\bar{\Gamma}^\rho}_{\mu\beta}{N^\mu}_{\sigma\alpha}\right) \label{sdef}
	\end{equation}
	
	The Ricci tensor and Ricci scalar are defined in the same manner as the traces of the Riemann curvature tensor,
	\begin{align}
		{R}_{\sigma\beta} &= g^{\rho\alpha}{R}_{\rho\sigma\alpha\beta} =  {{R}^\alpha}_{\sigma\alpha\beta} \label{riccitenma}\\
		{R} &= g^{\sigma\beta}{R}_{\sigma\beta} = {{R}^{\alpha\beta}}_{\alpha\beta} \label{ricciscalarma}
	\end{align}
	{Besides the usual Ricci tensor above, there exist two more independent traces of the Riemann curvature tensor,
		\begin{align}
			{R}_{\alpha\beta} &= g^{\rho\sigma}{R}_{\rho\sigma\alpha\beta} =  {{R}^\sigma}_{\sigma\alpha\beta} \label{riccitenma2}\\
			{R}_{\rho\beta} &= g^{\sigma\alpha}{R}_{\rho\sigma\alpha\beta} =  {{R_\rho}^\alpha}\ \!_{\alpha\beta} \label{riccitenma3}
		\end{align}
		These are called the homothetic tensor and co-Ricci tensor, respectively. Although there are three independent traces of the Riemann tensor, the Ricci scalar is uniquely defined since the trace of the homothetic tensor vanishes and that of the co-Ricci tensor is simply $ -R $.  }
	
	\section{Kinematics in Metric-Affine Geometry} \label{sec-makin}
	The geometric setup to analyse kinematics of curves in a metric-affine geometry is similar to that in Riemannian geometry. We will start with a congruence of curves with tangent vector, $ u^\alpha = \frac{\partial x^\alpha}{\partial t} $. However, one of the effects of the non-metricity of the geometry is that the {length of vectors is not preserved under parallel transportation}. Therefore, the norm of $ u^\alpha $ cannot be normalised like in equation \eqref{vecnorm}. Instead, now we will have,
	\begin{equation}\label{vecnormma}
		u^\alpha u_\alpha = \epsilon l^2
	\end{equation}
	where, $ l \equiv l(x^\alpha) $. The curves are called timelike, null, and spacelike for $ \epsilon = -1,0,+1 $, respectively. {One should also note that the non-metricity can also change the direction of the vectors.}
	
	Defining a deviation vector in the same fashion as for Riemannian geometry, one can check that equation \eqref{lieder1} still holds. However, equation \eqref{deviation} gets modified to,
	\begin{equation}\label{deviationma}
		\Df_u \xi^\alpha = u^\beta \nabla_\beta \xi^\alpha = \xi^\beta\left(\nabla_\beta u^\alpha + 2{T^\alpha}_{\beta\sigma}u^\sigma\right)
	\end{equation}
	To recognise the quantities that prevent the deviation vector from being orthogonal to the congruence, we calculate,
	\begin{equation}\label{devorthma}
		\Df_u \left(\xi^\alpha u_\alpha\right) = \xi^\alpha a_\alpha - 2T_{\beta\sigma\alpha}\xi^\alpha u^\beta u^\sigma + \hf L_\alpha \xi^\alpha - \hf Q_{\alpha\beta\sigma}\xi^\alpha u^\beta u^\sigma
	\end{equation}
	where, we define,
	\begin{equation}\label{veclengthma}
		L_\beta = \nabla_\beta \left(-u^\alpha u_\alpha\right) = \partial_\beta\left(-u^\alpha u_\alpha\right) = \bar{\nabla}_\beta \left(-u^\alpha u_\alpha\right)
	\end{equation}
	and,
	\begin{equation}\label{acclma1}
		a_\alpha = \Df_u u_\alpha\;  ; \quad a^\alpha = g^{\alpha\beta} a_\beta
	\end{equation}
	Therefore, in addition to the path acceleration of the curves, both the non-Riemannian variables and the non-constant length of the vectors contribute to the longitudinal component of the deviation vector.
	
	It is important to note here that due to the non-metricity of the geometry, raising and lowering indices is not trivial any longer. That is,
	\begin{equation}\label{raislow}
		{\Df_u u^\alpha \ne g^{\alpha\beta}\Df_u u_\beta}
	\end{equation}
	Therefore, we define another acceleration,
	\begin{equation}\label{acclma2}
		A^\alpha = \Df_u u^\alpha \ne a^\alpha\; ; \quad A_\alpha = g_{\alpha\beta}A^\beta \ne a_\alpha
	\end{equation}
	The two accelerations are related through the non-metricity tensor,
	\begin{equation}\label{acclma3}
		a_\alpha = A_\alpha + Q_{\rho\sigma\alpha}u^\rho u^\sigma
	\end{equation}
	Equation \eqref{devorthma} not being a constant even for $ a_\alpha = 0 $ (like in the case of Riemannian geometry) is a consequence of the evolution tensor not being orthogonal to the tangent vector which in turn is due to the presence of torsion and non-metricity. Therefore, to derive the kinematics of the congruence, we will have to derive the transverse component of the evolution tensor. 
	
	\subsection{Geodesics and Autoparallels}
	At this point, we would like to establish some terminology for differentiating between curves with zero `acceleration'. In a metric affine geometry, a given curve has two types of acceleration: $ A^\alpha = \Df_u u^\alpha $  and $ a_\alpha = \Df_u u_\alpha $. We will refer to the former as `path'-acceleration and the latter as `hyper'-acceleration \cite{ios}. Furthermore, we will refer to the acceleration defined using the Riemannian part of the directional derivative ($ \bar{a}^\alpha = \bar{\Df}_u u^\alpha $, $ \bar{a}_\alpha = g_{\alpha\sigma}\bar{\Df}_u u^\sigma $) as simply `Riemannian-acceleration'. These three quantities are related through equation \eqref{acclma3} and, additionally, the following relations,
	\begin{align}
		A^\alpha &= \bar{a}^\alpha + {N^\alpha}_{\beta\sigma}u^\beta u^\sigma\; \Rightarrow A_\alpha = \bar{a}_\alpha + N_{\alpha\beta\sigma}u^\beta u^\sigma \label{acclma4}\\
		a_\alpha &= \bar{a}_\alpha - N_{\beta\alpha\sigma}u^\beta u^\sigma \; \Rightarrow a^\alpha = \bar{a}^\alpha - {N_\beta}^\alpha\! _\sigma u^\beta u^\sigma \label{acclma5}
	\end{align}
	Then, we can classify the curves depending on which acceleration is taken to be zero. We will term the curves with zero Riemannian acceleration ($ \boldsymbol{\bar{a}} = 0 $) as `r-autoparallels'. On the other hand, we will call the curves with zero path-acceleration ($ \boldsymbol{A} = 0 $) as `nr-autoparallels'. {Further, geodesics are defined as the curves that extremise the distance between two points. In Riemannian geometries, the r-autoparallels are also geodesics.} However, neither r-autoparallels nor nr-autoparallels are necessarily geodesic curves in metric-affine geometry {(or in any non-Riemannian geometry with a non-zero torsion or a non-zero non-metricity or both)}. Therefore, it is important to distinguish between the two.
	
	Moreover, using equation \eqref{veclengthma}, we can write,
	\begin{equation}\label{fixlenvecma}
		u^\alpha\bar{a}_\alpha = -\hf u^\alpha L_\alpha = -\hf u_\alpha L^\alpha =  u_\alpha\bar{a}^\alpha 
	\end{equation} 
	Therefore, for r-autoparallels ($ \boldsymbol{\bar{a}} = 0 $), we see that $ u^\alpha L_\alpha = \frac{\df}{\df t}l^2 = 0 $, which means that the vectors are of constant length. Therefore, in a metric-affine geometry, taking curves to be r-autoparallels automatically means that their tangent vectors will have fixed length under parallel transport\footnote{The converse, however, need not be true. The integral curves of fixed length vectors are not necessarily r-autoparallels. Moreover, there also exist particular forms of the non-metricity tensor that allow for vectors to be of fixed length \cite{ios,iosth}. }. Here on, the term `fixed length' would mean $ L_\alpha = 0 $.
	
	\subsection{Deviation Equation}
	The deviation equation in the case of Riemannian geometry gave us an interpretation for the curvature which was that it prevents two initially parallel curves from remaining parallel. Therefore, it is of interest to check whether the additional features of a metric-affine geometry, namely, torsion and non-metricity, also contribute to this or not. To do this, we take a directional derivative of equation \eqref{deviationma} along the congruence to find,
	\begin{equation}\label{deviationma2}
		\Df^2_u \xi^\alpha = {R^\alpha}_{\beta\rho\sigma}u^\beta u^\rho \xi^\sigma + \xi^\sigma\nabla_\sigma A^\alpha - 2{T^\alpha}_{\beta\sigma}A^\beta\xi^\sigma \\- 2{T^\alpha}_{\beta\sigma}u^\beta\Df_u\xi^\sigma - 2\Df_u{T^\alpha}_{\beta\sigma}u^\beta\xi^\sigma 
	\end{equation}
	The above expression matches with the one derived in \cite{iosth}. It tells us that, in addition to the Riemann curvature tensor, both the torsion and non-metricity\footnote{It is important to note that even if non-metricity does not show up explicitly in equation \eqref{deviationma2} (and also in equations \eqref{ricciidenma} and \eqref{deviationma}), it does affect the acceleration of the deviation vector implicitly through the curvature tensor and covariant derivatives.} contribute to the acceleration of the deviation vector. That is, even with the Riemann curvature being zero, two initially parallel curves will not always remain parallel in a metric-affine geometry. In other words, even with a zero Riemann tensor, we are still in non-Euclidean space since the two other non-zero geometrical quantities, torsion and non-metricity, contribute to an effective curvature.
	
	Assuming nr-autoparallel motion, the deviation equation reduces to,
	\begin{equation}\label{deviationma3}
		\Df^2_u \xi^\alpha = {R^\alpha}_{\beta\rho\sigma}u^\beta u^\rho \xi^\sigma - 2{T^\alpha}_{\beta\sigma}u^\beta\Df_u\xi^\sigma - 2\Df_u{T^\alpha}_{\beta\sigma}u^\beta\xi^\sigma 
	\end{equation}
	The above equation can regarded as the `nr-autoparallel deviation' equation. Similarly, assuming the curves to be r-autoparallels and then using equation \eqref{acclma4}, the r-autoparallel deviation equation now becomes,
	\begin{multline}\label{deviationma4}
		\Df^2_u \xi^\alpha = {R^\alpha}_{\beta\rho\sigma}u^\beta u^\rho \xi^\sigma + \xi^\sigma\nabla_\sigma \left({N^\alpha}_{\beta\rho}u^\beta u^\rho\right) - 2{T^\alpha}_{\beta\sigma}{N^\beta}_{\rho\epsilon}u^\rho u^\epsilon\xi^\sigma \\- 2{T^\alpha}_{\beta\sigma}u^\beta\Df_u\xi^\sigma - 2\Df_u{T^\alpha}_{\beta\sigma}u^\beta\xi^\sigma 
	\end{multline}
	The contribution of the non-Riemannian variables to the relative acceleration between curves is more readily seen when these curves are taken to be r-autoparallels.
	
	\subsection{Kinematics of Timelike Curves}
	In this section, we derive the irreducible kinematics of a congruence of timelike curves ($ u^\alpha u_\alpha = -l^2 \ ; \; l(x^\alpha)\ne 0$). As in the Riemannian case, let us begin by rewriting equation \eqref{deviationma} as,
	\begin{equation}\label{evolma1}
		\Df_u\xi^\alpha = {B^\alpha}_\beta\xi^\beta
	\end{equation}
	where, we have,
	\begin{equation}\label{evolma2}
		{B^\alpha}_\beta = \nabla_\beta u^\alpha + 2{T^\alpha}_{\beta\sigma}u^\sigma
	\end{equation}
	The above $ (1,1) $ tensor, $ \boldsymbol{B} $, is now the fractional rate of change of the deviation vector. This is our evolution tensor now and we will define define our kinematic variables with respect to (the transverse component of) this tensor in order for them to encapsulate the transverse properties of the congruence in the same manner as in Riemannian geometry. As mentioned earlier, due to the non-metricity of the geometry, raising and lowering the indices needs to be done carefully. For this reason, we will define the $ (0,2) $ and $ (2,0) $ evolution tensors as,
	\begin{align}
		B_{\alpha\beta} &= g_{\alpha\rho}{B^\rho}_\beta \ne \nabla_\beta u_\alpha + 2T_{\alpha\beta\sigma}u^\sigma \\
		B^{\alpha\beta} &= {B^\alpha}_\rho g^{\rho\beta}
	\end{align}
	Using the above definition and separating out the Riemannian part, we can write,
	\begin{equation}\label{evolma3}
		B_{\alpha\beta} = \bar{B}_{\alpha\beta} + N_{\alpha\beta\sigma}u^\sigma
	\end{equation}
	The effect of the torsion and non-metricity on the evolution of the deviation tensor is more explicitly seen when we lower the indices. Equation \eqref{evolma3} is the one that we will use to define the kinematic variables and in other calculations for the rest of the paper.
	
	As in the Riemannian case, we are interested in the transverse properties of the congruence again, and hence, we need to separate out the transverse and longitudinal parts of the evolution tensor. For this purpose, we define a transverse metric, which will now be given by \cite{ios},
	\begin{equation}\label{transmetricma}
		h_{\alpha\beta} = g_{\alpha\beta} + \frac{1}{l^2}u_\alpha u_\beta
	\end{equation} 
	where, using $ u^\alpha u_\alpha = -l^2 $, it is easy to check that, $ u^\alpha h_{\alpha\beta} = 0 = h_{\alpha\beta}u^\beta $. Also, it follows that, $ {h^\alpha}_\sigma {h^\sigma}_\beta = {h^\alpha}_\beta $ and $ h^{\alpha\beta}h_{\alpha\beta} = {h^\alpha}_\alpha = n-1 $.
	Then, the transverse component of the evolution tensor is given by,
	\begin{equation}\label{transevolma1}
		\tr B_{\alpha\beta} = {h^\rho}_\alpha {h^\sigma}_\beta B_{\rho\sigma} = B_{\alpha\beta} + \frac{1}{l^2}A_\alpha u_\beta + \frac{1}{l^4}u_\alpha u_\beta A_\rho u^\rho + \frac{1}{l^2}u_\alpha \left(L_\beta + N_{\rho\beta\sigma}u^\rho u^\sigma\right) 
	\end{equation}
	where, we have used, 
	\begin{align}
		u^\alpha B_{\alpha\beta} &= L_\beta + N_{\alpha\beta\sigma}u^\alpha u^\sigma \label{evolmacont1} \\
		B_{\alpha\beta}u^\beta &= \bar{a}_\alpha + N_{\alpha\beta\sigma}u^\beta u^\sigma = A_\alpha \label{evolmacont2} \\
		B_{\alpha\beta}u^\alpha u^\beta &= A_\alpha u^\alpha = \hf L_\beta u^\beta + N_{\alpha\beta\sigma}u^\alpha u^\beta u^\sigma \label{evolmacont3}
	\end{align}
	Further, using equations  \eqref{transevolein}, \eqref{evolma3}, and \eqref{evolmacont2}, we get,
	\begin{multline}\label{transevolma}
		\tr B_{\alpha\beta} = \tr \bar{B}_{\alpha\beta} + \frac{1}{l^2}u_\alpha L_\beta + \left(\frac{1-l^2}{l^2}\right)A_\alpha u_\beta  + \frac{1}{l^4}u_\alpha u_\beta A^\rho u_\rho + N_{\alpha\beta\sigma}u^\sigma \\+ N_{\alpha\rho\sigma}u_\beta u^\rho u^\sigma + \frac{1}{l^2} u_\alpha N_{\rho\beta\sigma}u^\rho u^\sigma 
	\end{multline}
	Using this, we can define the kinematic quantities as,
	\begin{align}\label{kinequantsma}
		\theta &= g^{\alpha\beta}\tr B_{\alpha\beta} = \bar{\theta} + \mathfrak{t}\\ \omega_{\alpha\beta} &= \tr B_{[\alpha\beta]} = \bar{\omega}_{\alpha\beta} + \mathfrak{w}_{\alpha\beta} \\
		\sigma_{\alpha\beta} &= \tr B_{(\alpha\beta)} - \frac{1}{{h^\rho}_\rho}\theta h_{\alpha\beta} = \bar{\sigma}_{\alpha\beta} + \mathfrak{s}_{\alpha\beta}
	\end{align}
	where, the Fraktur symbols capture the non-Riemannian contribution to the kinematic quantities, and are given by,
	\begin{equation}\label{expnr}
		\mathfrak{t} = \left(\frac{1-l^2}{l^2}\right) A_\alpha u^\alpha + {N^\alpha}_{\alpha\beta}u^\beta + N_{\alpha\beta\sigma}u^\alpha u^\beta u^\sigma 
	\end{equation} 
	\begin{equation}\label{rotnr}
		\mathfrak{w}_{\alpha\beta} = \frac{1}{l^2} u_{[\alpha}L_{\beta]} + \left(\frac{1-l^2}{l^2}\right) A_{[\alpha}u_{\beta]} + N_{[\alpha\beta]\sigma}u^\sigma + N_{[\alpha\un{\rho}\un{\sigma}}u_{\beta]}u^\rho u^\sigma + \frac{1}{l^2}u_{[\alpha}N_{\un{\rho}\beta]\sigma}u^\rho u^\sigma
	\end{equation}
	\begin{multline}\label{shearnr}
		\mathfrak{s}_{\alpha\beta} =  \frac{1}{l^2} u_{(\alpha}L_{\beta)} - \frac{1-l^2}{(n-1)l^2}\left(\bar{\theta}u_\alpha u_\beta + h_{\alpha\beta}A_\rho u^\rho\right) - \frac{1}{n-1}h_{\alpha\beta}{N^\rho}_{\rho\sigma}u^\sigma \\- \frac{1}{n-1}h_{\alpha\beta}N_{\rho\sigma\epsilon}u^\rho u^\sigma u^\epsilon + \frac{1}{l^4}u_\alpha u_\beta A_\rho u^\rho+ \left(\frac{1-l^2}{l^2}\right)A_{(\alpha}u_{\beta)} + N_{(\alpha\beta)\sigma}u^\sigma \\+ N_{(\alpha\un{\rho}\un{\sigma}}u_{\beta)}u^\rho u^\sigma + \frac{1}{l^2}u_{(\alpha}N_{\un{\rho}\beta)\sigma}u^\rho u^\sigma 
	\end{multline}
	Once we have these expressions, we can go ahead and write the evolution equations for these quantities,
	\begin{multline}\label{kinema1}
		\Df_u\theta = \bar{\Df}_u\bar{\theta} - \frac{1}{l^4}u^\beta L_\beta A_\alpha u^\alpha + u^\beta\Df_u {N^\alpha}_{\alpha\beta} + {N^\alpha}_{\alpha\beta}A^\beta + u^\alpha u^\beta u^\sigma \Df_u N_{\alpha\beta\sigma} \\+ \left(\frac{1-l^2}{l^2}\right)\left(u^\alpha\Df_u A_\alpha + A_\alpha A^\alpha\right) + N_{\alpha\beta\sigma}A^\alpha u^\beta u^\sigma + N_{\alpha\beta\sigma}u^\alpha A^\beta u^\sigma + N_{\alpha\beta\sigma}u^\alpha u^\beta A^\sigma
	\end{multline}
	where, the first term on the right hand side is given by equation \eqref{einkin1}. The above equation is the most general form of the Raychaudhuri equation to appear in the literature, at least in the form that it is presented here. A different form of this equation has been presented in \cite{ios}. Note that $ {N^\alpha}_{\alpha\beta} = -\hf {Q_{\beta\alpha}}^\alpha $ and $ N_{\alpha\beta\sigma}u^\alpha u^\beta u^\sigma = -\hf Q_{\alpha\beta\sigma}u^\alpha u^\beta u^\sigma $. Therefore, it is only the non-metricity that affects the expansion of the congruence while torsion does not have any effect. This is consistent with the findings in \cite{luz,hensh,dey2,dey1}. 
	
	The equations for rotation and shear follow, and are given by,
	{\small
		\begin{multline}\label{kinema2}
			\Df_u \omega_{\alpha\beta} = \bar{\Df}_u\bar{\omega}_{\alpha\beta} - {N^\sigma}_{\alpha\rho}\bar{\omega}_{\sigma\beta}u^\rho - \bar{\omega}_{\alpha\sigma}{N^\sigma}_{\beta\rho}u^\rho -\frac{1}{l^4}u^\rho L_\rho u_{[\alpha}L_{\beta]} + \frac{1}{l^2}a_{[\alpha}L_{\beta]} + \frac{1}{l^2}u_{[\alpha}\Df_uL_{\beta]} \\- \frac{1}{l^4}u^\rho L_\rho A_{[\alpha}L_{\beta]} + N_{[\alpha\beta]\sigma}A^\sigma + \left(\frac{1-l^2}{l^2}\right)\left(\Df_uA_{[\alpha}u_{\beta]} + A_{[\alpha}a_{\beta]}\right) + \Df_u N_{[\alpha\beta]\sigma}u^\sigma + \Df_u N_{[\alpha\un{\rho}\un{\sigma}}u_{\beta]}u^\rho u^\sigma \\+ N_{[\alpha\un{\rho}\un{\sigma}}a_{\beta]}u^\rho u^\sigma + N_{[\alpha\un{\rho}\un{\sigma}}u_{\beta]}A^\rho u^\sigma + N_{[\alpha\un{\rho}\un{\sigma}}u_{\beta]}u^\rho A^\sigma  - \frac{1}{l^4}u^\epsilon L_\epsilon u_{[\alpha}N_{\un{\rho}\beta]\sigma}u^\rho u^\sigma \\+ \frac{1}{l^2}\left( a_{[\alpha}N_{\un{\rho}\beta]\sigma}u^\rho u^\sigma + u_{[\alpha}\Df_uN_{\un{\rho}\beta]\sigma}u^\rho u^\sigma + u_{[\alpha}N_{\un{\rho}\beta]\sigma}A^\rho u^\sigma + u_{[\alpha}N_{\un{\rho}\beta]\sigma}u^\rho A^\sigma \right)
		\end{multline}
		\begin{multline}\label{kinema3}
			\Df_u \sigma_{\alpha\beta} = \bar{\Df}_u\bar{\sigma}_{\alpha\beta} - {N^\rho}_{\alpha\epsilon}\bar{\sigma}_{\rho\beta}u^\epsilon - \bar{\sigma}_{\alpha\rho}{N^\rho}_{\beta\epsilon}u^\epsilon - \frac{1}{l^4}u^\rho L_\rho u_{(\alpha}L_{\beta)} + \frac{1}{l^2}a_{(\alpha}L_{\beta)} + \frac{1}{l^2}u_{(\alpha}\Df_u L_{\beta)} \\+ \frac{1}{(n-1)l^4}u^\rho L_\rho\bar{\theta}u_\alpha u_\beta - \frac{1-l^2}{(n-1)l^2}\left(\bar{\Df}_u\bar{\theta}u_\alpha u_\beta + 2\bar{\theta}a_{(\alpha} u_{\beta)}\right) \\ - \frac{1}{n-1}\left( u^\epsilon Q_{\epsilon\alpha\beta} - \frac{1}{l^4}u^\epsilon L_\epsilon u_\alpha u_\beta + \frac{2}{l^2}a_{(\alpha}u_{\beta)} \right) \left({N^\rho}_{\rho\sigma}u^\sigma + A_\rho u^\rho\right) \\- \frac{h_{\alpha\beta}}{n-1} \left( \Df_u {N^\rho}_{\rho\sigma}u^\sigma + {N^\rho}_{\rho\sigma}A^\sigma + \Df_u A_\rho u^\rho + A^\rho A_\rho\right) - \frac{2}{l^6}u^\rho L_\rho u_\alpha u_\beta A_\rho u^\rho - \frac{1}{l^4}u^\rho L_\rho A_{(\alpha}u_{\beta)} \\+ \frac{1}{l^4}\left( 2a_{(\alpha} u_{\beta)} A_\rho u^\rho + u_\alpha u_\beta \Df_uA_\rho u^\rho + u_\alpha u_\beta A_\rho A^\rho\right) + \left(\frac{1-l^2}{l^2}\right)\left(\Df_uA_{(\alpha}u_{\beta)} + A_{(\alpha}a_{\beta)}\right) \\+ \Df_uN_{(\alpha\beta)\sigma}u^\sigma + N_{(\alpha\beta)\sigma}A^\sigma + \Df_u N_{(\alpha\un{\rho}\un{\sigma}}u_{\beta)}u^\rho u^\sigma + N_{(\alpha\un{\rho}\un{\sigma}}a_{\beta)}u^\rho u^\sigma \\+ N_{(\alpha\un{\rho}\un{\sigma}}u_{\beta)}A^\rho u^\sigma + N_{(\alpha\un{\rho}\un{\sigma}}u_{\beta)}u^\rho A^\sigma - \frac{1}{l^4}u^\epsilon L_\epsilon u_{(\alpha}N_{\un{\rho}\beta)\sigma}u^\rho u^\sigma \\+ \frac{1}{l^2}\left( a_{(\alpha}N_{\un{\rho}\beta)\sigma}u^\rho u^\sigma + u_{(\alpha}\Df_uN_{\un{\rho}\beta)\sigma}u^\rho u^\sigma + u_{(\alpha}N_{\un{\rho}\beta)\sigma}A^\rho u^\sigma + u_{(\alpha}N_{\un{\rho}\beta)\sigma}u^\rho A^\sigma\right)
	\end{multline}}
	where the terms with a `bar' are given by equations \eqref{einkin1}, \eqref{einkin2}, and \eqref{einkin3}. The above two equations (kinematic equations for the rotation and shear) are appearing in the literature for the first time (in their current form). A different form of only the equation for rotation has been presented in \cite{iosth}.
	
	\subsubsection{Expansion of Timelike Curves and $ (n-1) $-Volume} \label{subsec-fracvol}
	In Riemannian geometry, the expansion scalar is exactly equal to the fractional rate of change of the cross sectional  $ (n-1) $-volume of the congruence. A formal proof of this can be found in \cite{poisson}. We will follow the same steps, as presented in \cite{poisson}, here to check how the expansion scalar of the metric-affine geometry is related to the $ (n-1) $-volume of the congruence. 
	
	To introduce the notion of cross sectional volume, we select a particular point, $ P $, corresponding to a parameter value, $ t_P $, on any given curve in the congruence. Then, the cross section around this curve at point, $ P $, can be constructed by taking a set of points, $ P^\prime $, on the neighbouring curves which have the same parameter value, $ t_P $. Let us call this set of points $\Sigma_P$. Then, we want to compare $ \Sigma_P $ to another such set of points, $ \Sigma_Q $, at some other point, $ Q $, on the curve where the parameter value is $ t_Q $. To construct a metric on $ \Sigma_P $, we introduce a coordinate system, $ y^a\ (a = 1,2,\dots, n-1) $. Since each point, $ P^\prime $, on the cross section is also a point on the neighbouring curve, this coordinate system can also be used as a label for these neighbouring curves. Then, identifying all the points on $ \Sigma_Q $ by the curves passing through it, we automatically get a coordinate system on $ \Sigma_Q $ as well. Then, the vectors, $ n^\alpha_a = \frac{\partial x^\alpha}{\partial y^a} $, are tangent to the cross section. 
	
	Now, for a distance between two points on the cross section, we have $ \df t = 0 $, and hence, we can write,
	\begin{align*}
		\df s^2 &= g_{\alpha\beta}\df x^\alpha \df x^\beta = g_{\alpha\beta} \frac{\partial x^\alpha}{\partial y^a}\df y^a \frac{\partial x^\beta}{\partial y^b}\df y^b \\
		\Rightarrow \df s^2 &= g_{\alpha\beta} n^\alpha_a n^\beta_b \df y^a \df y^b
	\end{align*}
	Therefore, the $ (n-1) $-tensor, $ h_{ab} = g_{\alpha\beta} n^\alpha_a n^\beta_b $, acts like the $ (n-1) $-metric of the cross section. Given this, the volume element on the section would be given by, $ \delta V = \sqrt{\det h_{ab}}\ \df^{n-1} y$. Since we set the coordinates, $ y^a $, up such that they remain the same along the curves (that is a curve retains its label), the change in the volume element as one goes from $ P $ to $ Q $ is only due to the change in $ \sqrt{\det h_{ab}} $. Therefore, one can write,
	\[ \frac{1}{V_{n-1}}\Df_u V_{n-1}  = \frac{1}{\sqrt{\det h_{ab}}}\Df_u \sqrt{\det h_{ab}} = \frac{1}{ 2h_{ab}}\Df_u h_{ab} \]
	
	A straightforward calculation gives,
	\begin{align}
		\Df_u h_{a b} &= \Df_u (g_{\alpha\beta}n^\alpha_a n^\beta_b) = \left(Q_{\rho\alpha\beta}u^\rho + 2B_{(\alpha\beta)} - 4 T_{(\alpha\beta)\rho}u^\rho\right)n^\alpha_a n^\beta_b \nonumber \\
		\Rightarrow \frac{1}{ 2h_{ab}}\Df_u h_{ab} &= \left(\hf Q_{\rho\alpha\beta}u^\rho + B_{(\alpha\beta)} - 2 T_{(\alpha\beta)\rho}u^\rho\right) g^{\alpha\beta} \nonumber\\
		\Rightarrow \frac{1}{ 2h_{ab}}\Df_u h_{ab} &= \hf {Q_{\rho \alpha}}^\alpha u^\rho + g^{\alpha\beta}\tr B_{\alpha\beta} - \frac{1}{l^2}A_\alpha u^\alpha - 2{T^\alpha}_{\alpha\rho}u^\rho \nonumber\\
		\Rightarrow \frac{1}{V_{n-1}}\Df_u V_{n-1}  &= \theta - \frac{1}{l^2}A_\alpha u^\alpha - 2{N^\alpha}_{\alpha\rho}u^\rho + {N^\alpha}_{\rho\alpha}u^\rho \label{fracvolchange}
	\end{align}
	Therefore, the expansion scalar of the metric-affine geometry is related to the fractional rate of change of the cross sectional $ (n-1) $-volume of the congruence through a scalar function, $ \Phi $, that depends on the distortion tensor and the path-acceleration of the curves,
	\begin{equation}\label{fracvolchange1}
		\theta + \Phi = \frac{1}{V_{n-1}}\Df_u V_{n-1}
	\end{equation}
	where, we have defined, $ \Phi = {N^\alpha}_{\rho\alpha}u^\rho - 2{N^\alpha}_{\alpha\rho}u^\rho - \frac{1}{l^2}A_\alpha u^\alpha $. We see that the deviation from the Riemannian meaning of expansion is caused only by the projection of the vectors associated with the distortion tensor along the congruence. Assuming nr-autoparallel curves ($ \boldsymbol{A} = 0 $), the scalar $ \Phi $ will be zero if $ {Q_{\rho \alpha}}^\alpha = 4{T^\alpha}_{\alpha\rho} $. 
	
	Using equation \eqref{expnr} in equation \eqref{fracvolchange}, we can write,
	\begin{equation}\label{fracvolchange2}
		\bar{\theta} - u^\alpha \bar{a}_\alpha - 2T_\alpha u^\alpha = \frac{1}{V_{n-1}}\Df_u V_{n-1}
	\end{equation}
	where, we have defined, $ T_\alpha = {T^\rho}_{\rho\alpha} $. Considering r-autoparallel curves, ($ \boldsymbol{\bar{a}} = 0 $), we see that the fractional rate of change of the volume will be equal to the Riemannian part of the expansion scalar if the projection, $ u^\alpha T_\alpha  = 0 $, of the vector associated with only the torsion tensor is taken to be zero.
	
	\subsubsection{Hypersurface Orthogonality of Timelike Curves}
	Consider a family of hypersurfaces described the condition, $ \zeta(x^\alpha) = \kappa $. The normal to these hypersurfaces would be given by, $ n_\alpha = \partial_\alpha\zeta $. Then, a congruence of curves is called hypersurface orthogonal if the tangent vector, $ u^\alpha $, is proportional to the normal, $ u_\alpha = -\lambda^2\partial_\alpha \zeta $, where, $\lambda$ is a scalar function. Considering the antisymmetric tensor, $ \bar{B}_{[\alpha\beta}u_{\sigma]} = \bar{\nabla}_{[\beta}u_{\alpha}u_{\sigma]} $, the orthogonality condition translates to this tensor being zero. This is known as the Frobenius theorem \cite{poisson}. In Riemannian geometry, this condition translates to the rotation tensor being zero. In this section, we will show that this condition is equivalent to a condition on the rotation tensor, $ \omega_{\alpha\beta} $, that relates it directly to the distortion tensor.
	
	The orthogonality condition in terms of the antisymmetric tensor above can be written as,
	\begin{equation}\label{frob1}
		\bar{B}_{[\alpha\beta}u_{\sigma]} = \frac{2}{3!}\left(\bar{B}_{[\alpha\beta]}u_{\sigma} + \bar{B}_{[\sigma\alpha]}u_{\beta} + \bar{B}_{[\beta\sigma]}u_{\alpha} \right) = 0 
	\end{equation} 
	Using equation \eqref{evolma3}, this becomes,
	\begin{equation*}
		\left(B_{[\alpha\beta]}u_\sigma + B_{[\sigma\alpha]}u_\beta + B_{[\beta\sigma]}u_\alpha \right) - \left(N_{[\alpha\beta]\rho}u_\sigma + N_{[\sigma\alpha]\rho}u_\beta + N_{[\beta\sigma]\rho}u_\alpha \right)u^\rho = 0
	\end{equation*} 
	Multiplying both sides by $ u^\sigma $, we get,
	\begin{equation*}
		\left(-l^2B_{[\alpha\beta]} + B_{[\sigma\alpha]}u_\beta u^\sigma + B_{[\beta\sigma]}u_\alpha u^\sigma \right) = \left(-l^2N_{[\alpha\beta]\rho} + N_{[\sigma\alpha]\rho}u_\beta u^\sigma + N_{[\beta\sigma]\rho}u_\alpha u^\sigma \right)u^\rho
	\end{equation*}
	Then, using equation \eqref{transevolma1} and equations \eqref{evolmacont1} - \eqref{evolmacont3}, we can write,
	\begin{equation*}
		\omega_{\alpha\beta} = \tr B_{[\alpha\beta]} = B_{[\alpha\beta]} + \frac{1}{l^2} B_{[\alpha\un{\sigma}}u_{\beta]} u^\sigma + \frac{1}{l^2} u_{[\alpha}B_{\un{\sigma}\beta]}u^\sigma + \frac{1}{l^4} u_{[\alpha} u_{\beta]} B_{\rho\sigma}u^\rho u^\sigma
	\end{equation*}
	The last term on the right is identically zero, and the rest of the terms can be rearranged to get,
	\begin{equation*}
		\omega_{\alpha\beta} = B_{[\alpha\beta]} + \frac{1}{l^2} B_{[\alpha\sigma]}u_\beta u^\sigma + \frac{1}{l^2} B_{[\sigma\beta]}u_\alpha u^\sigma
	\end{equation*} 
	Using this in the orthogonality condition above, we can write,
	\begin{equation*}
		-l^2 \omega_{\alpha\beta} = \left(-l^2N_{[\alpha\beta]\rho} + N_{[\sigma\alpha]\rho}u_\beta u^\sigma + N_{[\beta\sigma]\rho}u_\alpha u^\sigma \right)u^\rho
	\end{equation*} 
	which gives us,
	\begin{equation}\label{orthcondrot}
		\omega_{\alpha\beta} = N_{[\alpha\beta]\rho} u^\rho + \frac{1}{l^2}N_{[\alpha\sigma]\rho}u_\beta u^\rho u^\sigma + \frac{1}{l^2}N_{[\sigma\beta]\rho}u_\alpha u^\rho u^\sigma
	\end{equation}
	The above equation serves as a condition on the rotation tensor for the congruence to be hypersurface orthogonal. In other words, for a hypersurface orthogonal congruence, the rotation tensor can be determined directly from the distortion tensor and the tangent vector.
	
	\subsection{Kinematics of Null Curves}
	In this section, we aim to derive similar kinematic equations as the last section, but for null curves. A curve is called null if its tangent vector is a null vector. That is, $ k^\alpha k_\alpha = 0 $, where, $ k^\alpha = \frac{\partial x^\alpha}{\partial t} $, is the tangent to the curve. Now, as we have mentioned before, in a general metric-affine geometry, the length of tangent vectors is not preserved under parallel transport. Therefore, the null nature of a curve may not be preserved at all points. However, we want to consider curves that are null everywhere. That is, if $ k^\alpha k_\alpha = 0 $ at a particular point on the curve, it should be zero everywhere. To arrange this, we will demand that the norm of the tangent vector remains zero everywhere along the curve. This translates to, $ k^\beta \nabla_\beta\left( k^\alpha k_\alpha \right) = 0 $ if $ k^\alpha k_\alpha = 0 $ at any point. This is akin to having a fixed length vector, but the condition only applies to vectors which are null at any given point. This would also mean that the path-acceleration of null curves is solely due to the non-Riemannian features of the geometry. Then, to make a clear distinction between timelike and null curves, one can define timelike curves as the curves whose tangent vectors follow, $ u^\alpha u_\alpha = -l^2 $, where, $ l(x^\alpha) \ne 0 $. In what follows, this condition will be implicitly assumed.	
	
	To derive the kinematics of a congruence of null curves, we will follow analogous steps as in the previous section. To avoid confusion, we will relabel the null tangent vector to be, $ k^\alpha k_\alpha = 0 $, and the evolution tensor in equation \eqref{evolma2} and equation \eqref{evolma3} will also be relabelled accordingly. Again, we are interested in transverse properties of the congruence and for this reason we need to find the transverse component of the evolution tensor. To do this, we need to define a transverse metric, which is now given by,
	\begin{equation}\label{transmetricnullma}
		h_{\alpha\beta} = g_{\alpha\beta} + \frac{2}{m^2} k_{(\alpha}X_{\beta)}
	\end{equation}
	where, $ \boldsymbol{X} $ is an auxiliary null vector, $ X^\alpha X_\alpha = 0 $, such that, $ k^\alpha X_\alpha = -m^2 $, where, $ m^2 \equiv m^2(x^\alpha) $. Using this, it is easy to check that, $ k^\alpha h_{\alpha\beta} = 0 = h_{\alpha\beta}k^\beta $. Also, it follows that, $ {h^\alpha}_\sigma {h^\sigma}_\beta = {h^\alpha}_\beta $ and $ h^{\alpha\beta}h_{\alpha\beta} = {h^\alpha}_\alpha = n-2 $.
	
	Using the transverse metric, we can find the transverse component of the evolution tensor as,
	\begin{multline}\label{transevolmanull1}
		\tr B_{\alpha\beta} = B_{\alpha\beta} + \frac{1}{m^2}\left(A_\alpha X_\beta + Z_\alpha k_\beta\right) + \frac{1}{m^2} \left( X_\alpha N_{\rho\beta\sigma}k^\rho k^\sigma + k_\alpha Y_\beta \right) \\+ \frac{1}{m^4}\left( X_\alpha X_\beta k^\rho A_\rho + X_\alpha k_\beta k^\rho Z_\rho + k_\alpha X_\beta X^\rho A_\rho + k_\alpha k_\beta X^\rho Z_\rho  \right)
	\end{multline}
	where, we have used,
	\begin{align}
		k^\alpha B_{\alpha\beta} &= N_{\alpha\beta\sigma}k^\alpha k^\sigma \label{evolmanullcont1}\\
		B_{\alpha\beta}k^\beta &= \bar{a}_\alpha + N_{\alpha\beta\sigma}k^\beta k^\sigma = A_\alpha \label{evolmanullcont2}\\
		B_{\alpha\beta}k^\alpha k^\beta &= k^\alpha A_\alpha = N_{\alpha\beta\sigma}k^\alpha k^\beta k^\sigma \label{evolmanullcont3}
	\end{align}
	and we also define,
	\begin{align}
		X^\alpha B_{\alpha\beta} &= Y_\beta \; \Rightarrow k^\beta Y_\beta = X^\alpha A_\alpha\\
		B_{\alpha\beta}X^\beta &= Z_\alpha \; \Rightarrow k^\alpha Z_\alpha = N_{\alpha\beta\sigma}k^\alpha X^\beta k^\sigma\\
		B_{\alpha\beta}X^\alpha X^\beta &= X^\alpha Z_\alpha
	\end{align}
	Then, using equations \eqref{transevolnull}, \eqref{evolma3}, and \eqref{evolmanullcont2}, we get,
	\begin{multline}\label{transevolmanll2}
		\tr B_{\alpha\beta} = \tr \bar{B}_{\alpha\beta} - A_\alpha X_\beta + N_{\alpha\rho\sigma}X_\beta k^\rho k^\sigma - Z_\alpha k_\beta + N_{\alpha\beta\sigma}k^\sigma + N_{\alpha\rho\sigma}k_\beta X^\rho k^\sigma \\ - k_\alpha Y_\beta + k_\alpha N_{\rho\beta\sigma}X^\rho k^\sigma - k_\alpha X_\beta X^\rho A_\rho + k_\alpha X_\beta N_{\rho\sigma\epsilon}X^\rho k^\sigma k^\epsilon - k_\alpha k_\beta X^\rho Z_\rho \\+ k_\alpha k_\beta N_{\rho\sigma\epsilon}X^\rho X^\sigma k^\epsilon + \frac{1}{m^2}\left(A_\alpha X_\beta + Z_\alpha k_\beta\right) + \frac{1}{m^2} \left( X_\alpha N_{\rho\beta\sigma}k^\rho k^\sigma + k_\alpha Y_\beta \right) \\+ \frac{1}{m^4}\left( X_\alpha X_\beta k^\rho A_\rho + X_\alpha k_\beta k^\rho Z_\rho + k_\alpha X_\beta X^\rho A_\rho + k_\alpha k_\beta X^\rho Z_\rho  \right)
	\end{multline}
	We relabel the kinematic quantities as, $ \Theta, \Omega $, and $ \Sigma $, such that we have,
	\begin{align}\label{kinequantsmanill}
		\Theta &= g^{\alpha\beta}\tr B_{\alpha\beta} = \bar{\Theta} + \mathfrak{T}\\ \Omega_{\alpha\beta} &= \tr B_{[\alpha\beta]} = \bar{\Omega}_{\alpha\beta} + \mathfrak{W}_{\alpha\beta} \\
		\Sigma_{\alpha\beta} &= \tr B_{(\alpha\beta)} - \frac{1}{{h^\rho}_\rho}\Theta h_{\alpha\beta} = \bar{\Sigma}_{\alpha\beta} + \mathfrak{S}_{\alpha\beta}
	\end{align}
	where, the Fraktur symbols capture the non-Riemannian contribution to the kinematic quantities, and are given by,
	\begin{equation}\label{expnrnull}
		\mathfrak{T} = \left(\frac{1 - m^2}{m}\right)^2 X^\alpha A_\alpha + \left(2 - m^2\right)N_{\alpha\beta\sigma}X^\alpha k^\beta k^\sigma\\ + \frac{1}{m^2} k^\alpha Z_\alpha + {N^\alpha}_{\alpha\sigma}k^\sigma
	\end{equation}
	\begin{multline}\label{rotnrnull}
		\mathfrak{W}_{\alpha\beta} = \left(\frac{1-m^2}{m^2}\right)\left(A_{[\alpha}k_{\beta]} + Z_{[\alpha}k_{\beta]} + k_{[\alpha}Y_{\beta]}\right) + N_{[\alpha\beta]\sigma}k^\sigma + k_{[\alpha}N_{\un{\rho}\beta]\sigma}X^\rho X^\sigma \\+ N_{[\alpha\un{\rho}\un{\sigma}}k_{\beta]}X^\rho k^\sigma + N_{[\alpha\un{\rho}\un{\sigma}}X_{\beta]}k^\rho k^\sigma + X_{[\alpha}k_{\beta]}X^\rho A_\rho + k_{[\alpha}X_{\beta]}N_{\rho\sigma\epsilon}X^\rho k^\sigma k^\epsilon \\+ \frac{1}{m^2} X_{[\alpha}N_{\un{\rho}\beta]\sigma}k^\rho k^\sigma + \frac{1}{m^4}\left(X_{[\alpha}k_{\beta]}k^\rho Z_\rho + k_{[\alpha}X_{\beta]}X^\rho A_\rho\right)
	\end{multline}
	\begin{multline}\label{shearnrnull}
		\mathfrak{S}_{\alpha\beta} = - \frac{2(1-m^2)}{(n-2)m^2}\bar{\Theta}k_{(\alpha}X_{\beta)} + \left(\frac{1-m^2}{m^2}\right)\left(A_{(\alpha}k_{\beta)} + Z_{(\alpha}k_{\beta)} + k_{(\alpha}Y_{\beta)}\right) \\+ \frac{1}{m^2}X_{(\alpha}N_{\un{\rho}\beta)\sigma}k^\rho k^\sigma + \frac{1-m^4}{m^4} \left(k_{(\alpha}X_{\beta)}X^\rho A_\rho + k_{(\alpha}k_{\beta)}X^\rho Z_\rho\right) \\+ \frac{1}{m^4}\left( X_{(\alpha}k_{\beta)}k^\rho Z_\rho + X_{(\alpha}X_{\beta)}k^\rho A_\rho \right) + k_{(\alpha}X_{\beta)}N_{\rho\sigma\epsilon}X^\rho k^\sigma k^\epsilon + k_{(\alpha}k_{\beta)}N_{\rho\sigma\epsilon}X^\rho X^\sigma k^\epsilon \\+ N_{(\alpha\un{\rho}\un{\sigma}}X_{\beta)}k^\rho k^\sigma + N_{(\alpha\un{\rho}\un{\sigma}}k_{\beta)}X^\rho k^\sigma + k_{(\alpha}N_{\un{\rho}\beta)\sigma}X^\rho X^\sigma  + N_{(\alpha\beta)\sigma}k^\sigma \\-\frac{1}{n-2}h_{\alpha\beta}\left[ \left(\frac{1 - m^2}{m}\right)^2 X^\rho A_\rho + \left(2 - m^2\right)N_{\rho\sigma\epsilon}X^\rho k^\sigma k^\epsilon + \frac{1}{m^2} k^\rho Z_\rho + {N^\rho}_{\rho\sigma}k^\sigma \right] 
	\end{multline}
	
	Once we note these expressions down, it is straightforward to write the kinematic equations,
	\begin{multline}\label{kinemanull1}
		\Df_k \Theta = \bar{\Df}_k\bar{\Theta} - \frac{(1-m^4)}{m^4} k^\rho M_\rho X^\alpha A_\alpha + \left(\frac{1 - m^2}{m}\right)^2 \left(A_\alpha\Df_k X^\alpha + X^\alpha \Df_k A_\alpha\right) \\ - k^\rho M_\rho N_{\alpha\beta\sigma}X^\alpha k^\beta k^\sigma  - \frac{1}{m^4}k^\rho A_\rho k^\alpha Z_\alpha + \frac{1}{m^2}\left(A^\alpha Z_\alpha + k^\alpha\Df_kZ_\alpha\right) + \Df_k{N^\alpha}_{\alpha\sigma}k^\sigma + {N^\alpha}_{\alpha\sigma}A^\sigma \\ + (2-m^2)\left(X^\alpha k^\beta k^\sigma\Df_kN_{\alpha\beta\sigma} + N_{\alpha\beta\sigma}\Df_kX^\alpha k^\beta k^\sigma + N_{\alpha\beta\sigma}X^\alpha A^\beta k^\sigma + N_{\alpha\beta\sigma}X^\alpha k^\beta A^\sigma \right)
	\end{multline}
	where, we have defined, $ M_\beta = \nabla_\beta\left(-k^\alpha X_\alpha\right) = \bar{\nabla}_\beta\left(-k^\alpha X_\alpha\right) = \partial_\beta\left(-k^\alpha X_\alpha\right)$. The above equation is the most general form of the null Raychaudhuri equation (or Sachs optical equation) to appear in the literature. In GR, the expansion of congruence of null geodesics is directly related to the angular diameter distances, and hence, the above equation can be used to find an equation governing the cosmological distance measures in metric-affine theories. The first term on the right hand side is given by equation \eqref{einkinnull1}. Unlike the case of timelike curves, now both the non-metricity and torsion are affecting the expansion of null curves now. However, demanding that vectors have fixed lengths (or taking the Riemannian acceleration to be zero) makes the contribution from torsion vanish. Since in geometries where metricity holds, the vectors are by default of fixed lengths, this is again consistent with \cite{luz,hensh,dey1,dey2,dey3}. It is worth noting that a non-zero non-metricity does allow for the length of vectors to be preserved under parallel transport if we impose certain conditions on the form of the non-metricity \cite{ios,iosth}.
	
	The equations for rotation and shear are then given by,
	{\footnotesize
		\begin{multline}\label{kinemanull2}
			\Df_k\Omega = \bar{\Df}_k\bar{\Omega}  - {N^\sigma}_{\alpha\rho}\bar{\Omega}_{\sigma\beta}u^\rho - \bar{\Omega}_{\alpha\sigma}{N^\sigma}_{\beta\rho}u^\rho + \Df_kN_{[\alpha\beta]\sigma}k^\sigma + N_{[\alpha\un{\rho}\un{\sigma}}a_{\beta]}X^\rho k^\sigma - \frac{2}{m^6}k^\epsilon M_\epsilon \left(X_{[\alpha}k_{\beta]}k^\rho Z_\rho + k_{[\alpha}X_{\beta]}X^\rho A_\rho\right)\\+ a_{[\alpha}N_{\un{\rho}\beta]\sigma}X^\rho X^\sigma + k_{[\alpha}\Df_kN_{\un{\rho}\beta]\sigma}X^\rho X^\sigma + k_{[\alpha}N_{\un{\rho}\beta]\sigma}\Df_kX^\rho X^\sigma + k_{[\alpha}N_{\un{\rho}\beta]\sigma}X^\rho \Df_kX^\sigma + \Df_kN_{[\alpha\un{\rho}\un{\sigma}}k_{\beta]}X^\rho k^\sigma  + N_{[\alpha\beta]\sigma}A^\sigma \\+ N_{[\alpha\un{\rho}\un{\sigma}}k_{\beta]}\Df_kX^\rho k^\sigma + N_{[\alpha\un{\rho}\un{\sigma}}k_{\beta]}X^\rho A^\sigma + \Df_kN_{[\alpha\un{\rho}\un{\sigma}}X_{\beta]}k^\rho k^\sigma + N_{[\alpha\un{\rho}\un{\sigma}}\Df_kX_{\beta]}k^\rho k^\sigma + N_{[\alpha\un{\rho}\un{\sigma}}X_{\beta]}A^\rho k^\sigma + N_{[\alpha\un{\rho}\un{\sigma}}X_{\beta]}k^\rho A^\sigma \\+ \Df_kX_{[\alpha}k_{\beta]}X^\rho A_\rho + X_{[\alpha}k_{\beta]}\Df_kX^\rho A_\rho 
			+ X_{[\alpha}k_{\beta]}X^\rho \Df_kA_\rho + a_{[\alpha}X_{\beta]}N_{\rho\sigma\epsilon}X^\rho k^\sigma k^\epsilon + k_{[\alpha}\Df_kX_{\beta]}N_{\rho\sigma\epsilon}X^\rho k^\sigma k^\epsilon \\+ k_{[\alpha}X_{\beta]}\Df_kN_{\rho\sigma\epsilon}X^\rho k^\sigma k^\epsilon 
			+ k_{[\alpha}X_{\beta]}N_{\rho\sigma\epsilon}\Df_kX^\rho k^\sigma k^\epsilon + k_{[\alpha}X_{\beta]}N_{\rho\sigma\epsilon}X^\rho A^\sigma k^\epsilon + k_{[\alpha}X_{\beta]}N_{\rho\sigma\epsilon}X^\rho k^\sigma A^\epsilon + X_{[\alpha}a_{\beta]}X^\rho A_\rho\\ - \frac{1}{m^4}k^\epsilon M_\epsilon X_{[\alpha}N_{\un{\rho}\beta]\sigma}k^\rho k^\sigma 
			+ \frac{1}{m^2}\left( \Df_kX_{[\alpha}N_{\un{\rho}\beta]\sigma}k^\rho k^\sigma + X_{[\alpha}\Df_kN_{\un{\rho}\beta]\sigma}k^\rho k^\sigma
			+ X_{[\alpha}N_{\un{\rho}\beta]\sigma}A^\rho k^\sigma + X_{[\alpha}N_{\un{\rho}\beta]\sigma}k^\rho A^\sigma\right) \\ - \frac{1}{m^4}\left(A_{[\alpha}k_{\beta]} + Z_{[\alpha}k_{\beta]} + k_{[\alpha}Y_{\beta]}\right) + \frac{1-m^2}{m^2}\left(a_{[\alpha}A_{\beta]} + Z_{[\alpha}a_{\beta]} + a_{[\alpha}Y_{\beta]} - k_{[\alpha}\Df_kA_{\beta]} - k_{[\alpha}\Df_kZ_{\beta]} - k_{[\alpha}\Df_kY_{\beta]}\right)\\ + \frac{1}{m^4}\left(\Df_k X_{[\alpha}k_{\beta]}k^\rho Z_\rho + X_{[\alpha}a_{\beta]}k^\rho Z_\rho + X_{[\alpha}k_{\beta]}A^\rho Z_\rho + X_{[\alpha}k_{\beta]}k^\rho \Df_kZ_\rho + a_{[\alpha}X_{\beta]}X^\rho A_\rho  + k_{[\alpha}\Df_kX_{\beta]}X^\rho A_\rho \right. \\ \left.+ k_{[\alpha}X_{\beta]}\Df_kX^\rho A_\rho  + k_{[\alpha}X_{\beta]}X^\rho \Df_kA_\rho \right)
		\end{multline}
		
		\begin{multline}\label{kinemanull3}
			\Df_k \Sigma_{\alpha\beta} = \bar{\Df}_k\bar{\Sigma}  - 		{N^\sigma}_{\alpha\rho}\bar{\Sigma}_{\sigma\beta}u^\rho - \bar{\Sigma}_{\alpha\sigma}{N^\sigma}_{\beta\rho}u^\rho + \frac{2}{(n-2)m^4}k^\epsilon M_\epsilon \bar{\Theta}k_{(\alpha}X_{\beta)} + k_{(\alpha}N_{\un{\rho}\beta)\sigma}X^\rho \Df_kX^\sigma - \frac{k^\epsilon M_\epsilon}{m^4} X_{(\alpha}N_{\un{\rho}\beta)\sigma}k^\rho k^\sigma \\
			- \frac{2(1-m^2)}{(n-2)m^2}\left(\bar{\Df}_k\bar{\Theta}k_{(\alpha}X_{\beta)} + \bar{\Theta}a_{(\alpha}X_{\beta)} + \bar{\Theta}k_{(\alpha}\Df_kX_{\beta)}\right) 
			- \frac{1}{m^4}k^\epsilon M_\epsilon \left(A_{(\alpha}k_{\beta)} + Z_{(\alpha}k_{\beta)} + k_{(\alpha}Y_{\beta)}\right) + N_{(\alpha\beta)\sigma}A^\sigma \\
			+ \frac{1-m^2}{m^2}\left(\Df_kA_{(\alpha}k_{\beta)} + A_{(\alpha}a_{\beta)} + \Df_kZ_{(\alpha}k_{\beta)} + Z_{(\alpha}a_{\beta)} + a_{(\alpha}Y_{\beta)} + k_{(\alpha}\Df_kY_{\beta)}\right)+ \frac{2(1-m^4)}{m^4}a_{(\alpha}k_{\beta)}X^\rho Z_\rho  \\
			+\frac{1}{m^2}\left( \Df_kX_{(\alpha}N_{\un{\rho}\beta)\sigma}k^\rho k^\sigma  + X_{(\alpha}\Df_kN_{\un{\rho}\beta)\sigma}k^\rho k^\sigma + X_{(\alpha}N_{\un{\rho}\beta)\sigma}A^\rho k^\sigma + X_{(\alpha}N_{\un{\rho}\beta)\sigma}k^\rho A^\sigma\right) + \frac{2}{m^4} X_{(\alpha}\Df_kX_{\beta)}k^\rho A_\rho \\ + k_{(\alpha}\Df_kX_{\beta)}N_{\rho\sigma\epsilon}X^\rho k^\sigma k^\epsilon + a_{(\alpha}X_{\beta)}N_{\rho\sigma\epsilon}X^\rho k^\sigma k^\epsilon + k_{(\alpha}X_{\beta)}\Df_kN_{\rho\sigma\epsilon}X^\rho k^\sigma k^\epsilon+ k_{(\alpha}X_{\beta)}N_{\rho\sigma\epsilon}X^\rho A^\sigma k^\epsilon + k_{(\alpha}X_{\beta)}N_{\rho\sigma\epsilon}X^\rho k^\sigma A^\epsilon \\ + 2a_{(\alpha}k_{\beta)}N_{\rho\sigma\epsilon}X^\rho X^\sigma k^\epsilon + k_{(\alpha}k_{\beta)}\Df_kN_{\rho\sigma\epsilon}X^\rho X^\sigma k^\epsilon -\frac{2}{m^6} \left(k_{(\alpha}X_{\beta)}X^\rho A_\rho + k_{(\alpha}k_{\beta)}X^\rho Z_\rho + X_{(\alpha}k_{\beta)}k^\rho Z_\rho + X_{(\alpha}X_{\beta)}k^\rho A_\rho \right) \\ + \frac{1-m^4}{m^4}\left(a_{(\alpha}X_{\beta)}X^\rho A_\rho + k_{(\alpha}\Df_kX_{\beta)}X^\rho A_\rho + k_{(\alpha}X_{\beta)}\Df_kX^\rho A_\rho + k_{(\alpha}X_{\beta)}X^\rho \Df_kA_\rho \right)  + \Df_kN_{(\alpha\beta)\sigma}k^\sigma \\ - \frac{h_{\alpha\beta}}{n-2}\left[ \frac{1-m^4}{m^4} k^\rho M_\rho X^\alpha A_\alpha  - \frac{1}{m^4}k^\rho A_\rho k^\alpha Z_\alpha - k^\rho M_\rho N_{\alpha\beta\sigma}X^\alpha k^\beta k^\sigma + \Df_k{N^\alpha}_{\alpha\sigma}k^\sigma + {N^\alpha}_{\alpha\sigma}A^\sigma  + \frac{1}{m^2}\left(A^\alpha Z_\alpha + k^\alpha\Df_kZ_\alpha\right) \right. \\ \left. + (2-m^2)\left(X^\alpha k^\beta k^\sigma\Df_kN_{\alpha\beta\sigma} + N_{\alpha\beta\sigma}\Df_kX^\alpha k^\beta k^\sigma + N_{\alpha\beta\sigma}X^\alpha A^\beta k^\sigma + N_{\alpha\beta\sigma}X^\alpha k^\beta A^\sigma \right) + \left(\frac{1 - m^2}{m}\right)^2 \left(A_\alpha\Df_k X^\alpha + X^\alpha \Df_k A_\alpha\right) \right] \\ + \frac{1}{m^4}\left( \Df_kX_{(\alpha}k_{\beta)}k^\rho Z_\rho + X_{(\alpha}a_{\beta)}k^\rho Z_\rho + X_{(\alpha}k_{\beta)}A^\rho Z_\rho + X_{(\alpha}k_{\beta)}k^\rho \Df_kZ_\rho \right)  + k_{(\alpha}X_{\beta)}N_{\rho\sigma\epsilon}\Df_kX^\rho k^\sigma k^\epsilon
			+ k_{(\alpha}k_{\beta)}N_{\rho\sigma\epsilon}\Df_kX^\rho X^\sigma k^\epsilon\\ + k_{(\alpha}k_{\beta)}N_{\rho\sigma\epsilon}X^\rho\Df_k X^\sigma k^\epsilon + k_{(\alpha}k_{\beta)}N_{\rho\sigma\epsilon}X^\rho X^\sigma A^\epsilon + \Df_kN_{(\alpha\un{\rho}\un{\sigma}}X_{\beta)}k^\rho k^\sigma + N_{(\alpha\un{\rho}\un{\sigma}}\Df_kX_{\beta)}k^\rho k^\sigma + N_{(\alpha\un{\rho}\un{\sigma}}X_{\beta)}A^\rho k^\sigma + N_{(\alpha\un{\rho}\un{\sigma}}X_{\beta)}k^\rho A^\sigma \\+ \Df_kN_{(\alpha\un{\rho}\un{\sigma}}k_{\beta)}X^\rho k^\sigma + N_{(\alpha\un{\rho}\un{\sigma}}a_{\beta)}X^\rho k^\sigma + N_{(\alpha\un{\rho}\un{\sigma}}k_{\beta)}\Df_kX^\rho k^\sigma  + N_{(\alpha\un{\rho}\un{\sigma}}k_{\beta)}X^\rho A^\sigma + a_{(\alpha}N_{\un{\rho}\beta)\sigma}X^\rho X^\sigma + k_{(\alpha}\Df_kN_{\un{\rho}\beta)\sigma}X^\rho X^\sigma \\+ k_{(\alpha}N_{\un{\rho}\beta)\sigma}\Df_kX^\rho X^\sigma - \frac{1}{n-2}\left[k^\rho Q_{\rho\alpha\beta} - \frac{2}{m^4}k^\rho M_\rho k_{(\alpha}X_{\beta)} + \frac{2}{m^2}\left(a_{(\alpha}X_{\beta)} + k_{(\alpha}\Df_kX_{\beta)}\right)\right] \\	\times \left[ \left(\frac{1 - m^2}{m}\right)^2 X^\rho A_\rho + \left(2 - m^2\right)N_{\rho\sigma\epsilon}X^\rho k^\sigma k^\epsilon + \frac{1}{m^2} k^\rho Z_\rho + {N^\rho}_{\rho\sigma}k^\sigma \right] 
		\end{multline}
	}
	
	\subsubsection{Expansion of Null Curves and $ (n-2) $-Volume}
	In this section, we check whether the expansion scalar of the null curves is still the fractional rate of change of the cross sectional $ (n-2) $-volume of the congruence. The geometric setup in the null case is the same as for the timelike curves (refer to section \ref{subsec-fracvol} and also \cite{poisson}). We start by developing the notion of a cross sectional volume and define a $ (n-2) $-metric on cross section, $ h_{ab} = g_{\alpha\beta}n^\alpha_a n^\beta_b$. Then, following the same steps as in the case of timelike curves, the fractional rate of change of the $ (n-2) $-volume is given by,
	\begin{equation}\label{fracareachnage}
		\frac{1}{V_{n-2}}\Df_k V_{n-2} = \frac{1}{2h_{ab}}\Df_k h_{ab} = \Theta - 2{N^\alpha}_{\alpha\rho}k^\rho + {N^\alpha}_{\rho\alpha}k^\rho - \frac{1}{m^2}\left(A_\alpha X^\alpha + N_{\alpha\beta\sigma}k^\alpha X^\beta k^\sigma\right) 
	\end{equation}
	Therefore, contrary to Riemannian geometry, the expansion scalar in metric-affine geometry is not exactly equal to the fractional rate of change of the cross sectional $ (n-2) $-volume of the congruence, but rather is related to it through a scalar function that depends on the distortion tensor and the acceleration of the curves,
	\begin{equation}\label{fracareachange1}
		\Theta + \Psi = \frac{1}{V_{n-2}}\Df_k V_{n-2} 
	\end{equation}
	where, we have defined $ \Psi = {N^\alpha}_{\rho\alpha}k^\rho - 2{N^\alpha}_{\alpha\rho}k^\rho - \frac{1}{m^2}\left(A_\alpha X^\alpha + N_{\alpha\beta\sigma}k^\alpha X^\beta k^\sigma\right)  $. Here, we see that in addition to the projection of the distortion tensor along the curves, its projection along the auxiliary vector is also affecting the fractional rate of change of the cross-sectional volume. Assuming nr-autoparallel curves, the scalar $ \Psi $ will be zero if $ {N^\alpha}_{\rho\alpha}k^\rho = 2{N^\alpha}_{\alpha\rho}k^\rho + \frac{1}{m^2} N_{\alpha\beta\sigma}k^\alpha X^\beta k^\sigma $, which again translates to $ Q_{\sigma\beta\alpha} = 4T_{\alpha\beta\sigma} $.
	
	Using equation \eqref{expnrnull} in equation \eqref{fracareachnage}, this becomes,
	\begin{equation}\label{fracareachange2}
		\bar{\Theta} + (2-m^2)X^\alpha \bar{a}_\alpha - 2T_\alpha k^\alpha = \frac{1}{V_{n-2}}\Df_k V_{n-2} 
	\end{equation}
	where, we have again defined, $ T_\alpha = {T^\rho}_{\rho\alpha}  $. Assuming r-autoparallels, we see that it is only the projection of torsion along the curves that is affecting the fractional rate of change of volume. It will be equal to the Riemannian part of expansion if this projection is taken to be zero..
	
	\subsubsection{Hypersurface Orthogonality of Null Curves}
	Following the same arguments as for the case of timelike curves, the congruence being hypersurface orthogonal corresponds to the condition given by equation \eqref{frob1}. In the case of null curves, this does not translate to a direct condition on the rotation tensor. Rather, the rotation still remains complicated and is given by,
	{\small
		\begin{multline}\label{frobnull1}
			\Omega_{\alpha\beta} = N_{[\alpha\beta]\rho}k^\rho + \frac{X_{[\alpha} k_{\beta]}}{m^4}\left( N_{\rho\sigma\epsilon}k^\rho X^\sigma k^\epsilon - X^\alpha A_\alpha \right) \\+ \frac{1}{m^2}\left( k_\alpha N_{[\sigma\beta]\rho}X^\sigma k^\rho + N_{[\alpha\sigma]\rho}k_\beta X^\sigma k^\rho + A_{[\alpha}X_{\beta]}  N_{\sigma[\alpha\un{\rho}}X_{\beta]}k^\sigma k^\rho \right) 
	\end{multline}}
	Therefore, we see that for hypersurface orthogonal null congruences, the rotation tensor depends on both the distortion tensor and the acceleration of the curves. If we take the curves to be either nr-autoparallels or r-autoparallels, then the rotation tensor is completely determined by the distortion tensor, the tangent vector and the auxiliary vector.
	
	\section{Special Cases} \label{sec-spcase}
	The analysis in this paper has been quite general until this point. However, we want the results in this paper to be readily applicable to the diverse landscape of non-Riemannian theories of gravity. To this end, in this section, we will present the equations for expansion (the Raychaudhuri equation and the Sachs equation) in forms specific to the geometries employed in the respective gravity theories. To be able to distinguish between these cases, we will employ terminology given in table \ref{tb-magclass}. However, before we write the relevant kinematic equations, let us define the vectorial forms of the non-Riemannian variables that are often used in these theories.	
	
	\begin{table}[H]
		\caption{Different gravity theories as special cases of the metric-affine gravity.}
		\label{tb-magclass}
		\centering
		\begin{tabular}{@{}lllll}
			\hline
			\small\bf Geometry & \bf Gravity Theory & \bf Curvature & \bf Torsion & \bf Non-metricity \\
			\hline
			\multirow{5}{*}{Non-Riemannian}& Metric-affine & Non-zero & Non-zero & Non-zero \\
			
			&Einstein-Weyl-Cartan & Non-zero & Non-zero & Vectorial \\
			
			&Einstein-Cartan & Non-zero & Non-zero & Zero \\
			
			&Einstein-Weyl & Non-zero & Zero & Vectorial \\
			
			&TEGR & Zero & Non-zero & Zero \\
			
			&STEGR & Zero & Zero & Non-zero \\
			\hline
			Riemannian & Einstein & Non-zero & Zero & Zero \\
			\hline
			Flat & Newton & Zero & Zero & Zero \\
			\hline
		\end{tabular}
	\end{table}

	\subsection{Specific Forms of Torsion and Non-Metricity}
	\subsubsection{Vectorial Torsion} 
	Recall, that the torsion tensor, $ \boldsymbol{T} $, is defined as, $ {T^\rho}_{\alpha\beta} = {\Gamma^\rho}_{[\alpha\beta]} $. An associated vector can be defined as, $ T_\beta~=~g^{\rho\alpha} {T}_{\rho\alpha\beta}~=~{T^\alpha}_{\alpha\beta} $. {This vector part of the torsion tensor is responsible for r-autoparallels to differ from nr-autoparallels even with a vanishing the non-metricity. This is also the term that affects the fractional rate of change of the cross-sectional volume of the r-autoparallel congruences (equation \eqref{fracvolchange2}).} Using this, a simple form of torsion can be defined by constructing an antisymmetric tensor with help of the associated torsion vector. This is given by,
	\begin{equation}\label{torvec}
		{{\hat{T}}^\rho}\, _{\alpha\beta} = \frac{2}{n-1} {\delta^\rho}_{[\alpha}T_{\beta]}
	\end{equation}
	
	\subsubsection{Vectorial (or Weyl) Non-metricity}
	Recall, again, that the non-metricity, $ \boldsymbol{Q} $, is defined as, $ Q_{\rho\alpha\beta} = \nabla_\rho g_{\alpha\beta} $. Using this, one can define a vector associated with the non-metricity tensor as, $ Q_\rho = Q_{\rho\alpha\beta} g^{\alpha\beta} = {Q_{\rho\alpha}}^\alpha $. This is referred to as Weyl vector. Using this, a simple non-metricity tensor can be defined by constructing a symmetric tensor with the help of the associated vector. This is given by,
	\begin{equation}\label{nonmetvec}
		\hat{Q}_{\rho\alpha\beta} = \frac{1}{n} Q_\rho g_{\alpha\beta}
	\end{equation} 
	This is sometimes referred to as Weyl non-metricity.

	\subsection{Einstein-Weyl-Cartan Gravity}
	Taking the non-metricity to be vectorial as defined in equation \eqref{nonmetvec}, we find that, for timelike curves,
	\begin{equation}\label{ewcexp}
		\theta = \bar{\theta} + \frac{1-l^2}{l^2}\left(u^\alpha A_\alpha - \hf l^2 u^\alpha Q_\alpha\right)
	\end{equation}
	Therefore, it is the path-acceleration and the component of the Weyl vector along the tangent vector that affect the expansion. For nr-autoparallels, we have, $ \boldsymbol{A} = 0 $, and therefore, the only non-Riemannian contribution to expansion of the timelike congruence would be from the Weyl vector. For r-autoparallels ($ \boldsymbol{\bar{a}} = 0 $) on the other hand, all non-Riemannian contributions vanish. It is important to note that curves with zero Riemannian acceleration correspond to curves with tangent vectors of fixed length. Therefore, the vanishing of the non-Riemannian contributions can also be seen directly by putting $ l^2 = 1 $ in the above equation. The corresponding Raychaudhuri equation will be given by,
	\begin{multline}\label{rayewc}
		\Df_u \theta = \bar{\Df}_u \bar{\theta} - \frac{u^\rho L_\rho}{l^4} u^\alpha A_\alpha  + \frac{u^\rho L_\rho}{2l^2} u^\alpha Q_\alpha - \frac{1-l^2}{2} u^\alpha\Df_u Q_\alpha \\ + \frac{1-l^2}{l^2}\left(A^\alpha A_\alpha + u^\alpha\Df_u A_\alpha - \frac{l^2}{2}A^\alpha Q_\alpha\right) 
	\end{multline}
	
	Doing the same analysis for null congruences, we find,
	\begin{equation}\label{ewcexpnull}
		\Theta = \bar{\Theta} + \hf Q_\alpha k^\alpha + \frac{(1-m^2)^2}{m^2}\left(X^\alpha A_\alpha - 2T_{\beta\sigma\alpha}X^\alpha k^\beta k^\sigma + m^2 Q_\alpha k^\alpha\right)
	\end{equation}
	Now, for null congruences, there is contribution from both torsion and non-metricity even for nr-autoparallel curves. The effects from torsion vanish if  either we take $ \boldsymbol{\bar{a}} = 0 $ or if we take fixed length vectors ($ m^2 = 1 $). {Furthermore, if one takes the vectorial torsion (equation \eqref{torvec}), the expansion scalar becomes,
		\begin{equation}\label{ewcexpnullvector}
			\Theta = \bar{\Theta} + \hf Q_\alpha k^\alpha + \frac{(1-m^2)^2}{m^2}\left(X^\alpha A_\alpha - \frac{2 m^2}{3}T_\alpha k^\alpha + m^2 Q_\alpha k^\alpha\right)
		\end{equation}
		On the other hand, with the totally antisymmetric part of the torsion, $ T_{\beta\sigma\alpha} = T_{[\beta\sigma\alpha]} $, the torsion term in equation \eqref{ewcexpnull} vanishes.}
	
	The Sachs equation corresponding to \eqref{ewcexpnull} is be given by,
	\begin{multline}\label{sachsewcnull}
		\Df_k\Theta = \bar{\Df}_k\bar{\Theta} - \frac{1-m^4}{m^4}k^\rho M_\rho \left(X^\alpha A_\alpha - 2T_{\alpha\beta\sigma}k^\alpha k^\beta X^\sigma + m^2 Q_\alpha k^\alpha\right) + \hf \Df_k Q_\alpha k^\alpha \\+ \hf Q_\alpha A^\alpha + \frac{(1-m^2)^2}{m^2}\left(\Df_k X^\alpha A_\alpha + X^\alpha \Df_k A_\alpha - 2\Df_k T_{\alpha\beta\sigma}k^\alpha k^\beta X^\sigma - 2T_{\alpha\beta\sigma}A^\alpha k^\beta X^\sigma \right.\\ \left. - T_{\alpha\beta\sigma}k^\alpha A^\beta X^\sigma  - T_{\alpha\beta\sigma}k^\alpha k^\beta \Df_kX^\sigma + Q_\alpha M_\beta k^\alpha k^\beta + m^2\Df_k Q_\alpha k^\alpha + m^2 Q_\alpha A^\alpha  \right)
	\end{multline}
	{ With the vectorial torsion, this becomes,
		\begin{multline}\label{sachsewcnull2}
			\Df_k\Theta = \bar{\Df}_k\bar{\Theta} - \frac{1-m^4}{m^4}k^\rho M_\rho \left(X^\alpha A_\alpha - \frac{2 m^2}{3}T_\alpha k^\alpha + m^2 Q_\alpha k^\alpha\right) + \hf \Df_k Q_\alpha k^\alpha \\+ \hf Q_\alpha A^\alpha + \frac{(1-m^2)^2}{m^2}\left(\Df_k X^\alpha A_\alpha + X^\alpha \Df_k A_\alpha - \frac{2}{3}\left(T_\alpha M_\beta k^\alpha k^\beta + m^2\Df_k T_\alpha k^\alpha \right.\right.\\ \left.\left. + m^2 T_\alpha A^\alpha\right) + Q_\alpha M_\beta k^\alpha k^\beta + m^2\Df_k Q_\alpha k^\alpha + m^2 Q_\alpha A^\alpha  \right)
		\end{multline}
	}
	\subsection{Einstein-Cartan Gravity}
	Taking the non-metricity to be zero (which also means fixed length vectors) in the above equations makes the expansion scalar in both the cases simply equal to its Riemannian value. That is, $ \theta = \bar{\theta} $ and $ \Theta = \bar{\Theta} $. As mentioned earlier too, this is consistent with the findings in \cite{luz,hensh,dey2}.
	
	\subsection{Einstein-Weyl Gravity}
	Taking the torsion to be zero in equations \eqref{ewcexp} - \eqref{sachsewcnull} gives us the relevant equations for this case. Therefore, the equations for the expansion of timelike congruences remain the same. For null congruences, we will have,
	\begin{equation}\label{ewexpnull}
		\Theta = \bar{\Theta} + \frac{(1-m^2)^2}{m^2}\left(X^\alpha A_\alpha + m^2 Q_\alpha k^\alpha\right) + \hf Q_\alpha k^\alpha 
	\end{equation}
	and,
	\begin{multline}\label{sachsew}
		\Df_k\Theta = \bar{\Df}_k\bar{\Theta} - \frac{1-m^4}{m^4}k^\rho M_\rho \left(X^\alpha A_\alpha + m^2 Q_\alpha k^\alpha\right) + \hf \Df_k Q_\alpha k^\alpha + \hf Q_\alpha A^\alpha \\ + \frac{(1-m^2)^2}{m^2}\left(\Df_k X^\alpha A_\alpha + X^\alpha \Df_k A_\alpha  + Q_\alpha M_\beta k^\alpha k^\beta + m^2\Df_k Q_\alpha k^\alpha + m^2 Q_\alpha A^\alpha  \right)
	\end{multline}
	
	\subsection{TEGR}
	In this case, the expansion scalars again reduce to their Riemannian forms and so do their evolution equations. The kinematics in TEGR and Einstein-Cartan gravity may still be different from each other depending on the value of the Ricci tensor calculated with respect to the Levi-Civita connections that enters the kinematic equations.
	
	\subsection{STEGR}
	In this case, the expansion scalar for a timelike congruence and the Raychaudhuri equation remains the same as in the general metric-affine geometry (equations \eqref{expnr} and \eqref{kinema1}). However, for a null congruence, we will have,
	\begin{equation}\label{stegrexpnull}
		\Theta = \bar{\Theta} + \frac{(1-m^2)^2}{m^2} \left(X^\alpha A_\alpha - \hf Q_{\alpha\beta\sigma}X^\alpha k^\beta k^\sigma\right) - (2-m^2) Q_{\beta\sigma\alpha}X^\alpha k^\beta k^\sigma - \hf Q_\alpha k^\alpha
	\end{equation}
	Then, the corresponding Sachs equation will be given by,
	\begin{multline}\label{sachsstegr}
		\Df_k \Theta = \bar{\Df}_k\bar{\Theta} - \frac{1-m^4}{m^4} \left(X^\alpha A_\alpha - \hf Q_{\alpha\beta\sigma}X^\alpha k^\beta k^\sigma\right) - \hf Q_\alpha A^\alpha - \hf \Df_k Q_\alpha k^\alpha \\+ \frac{(1-m^2)^2}{m^2} \left(\Df_kX^\alpha A_\alpha + X^\alpha \Df_kA_\alpha \right) + k^\rho M_\rho Q_{\beta\sigma\alpha}X^\alpha k^\beta k^\sigma \\ - \frac{(1-m^2)^2}{2m^2} \left( \Df_kQ_{\alpha\beta\sigma}X^\alpha k^\beta k^\sigma + Q_{\alpha\beta\sigma}\Df_kX^\alpha k^\beta k^\sigma + Q_{\alpha\beta\sigma}X^\alpha A^\beta k^\sigma + Q_{\alpha\beta\sigma}X^\alpha k^\beta A^\sigma\right) \\ - (2-m^2)\left( \Df_kQ_{\beta\sigma\alpha}X^\alpha k^\beta k^\sigma + Q_{\beta\sigma\alpha}\Df_kX^\alpha k^\beta k^\sigma + Q_{\beta\sigma\alpha}X^\alpha A^\beta k^\sigma + Q_{\beta\sigma\alpha}X^\alpha k^\beta A^\sigma\right) 
	\end{multline}
	Note that the above expression simplifies greatly just by assuming fixed length vectors ($ m^2 = 1 $).

	\section{Cosmological Distances} \label{sec-cosdist}
	To be able to do cosmology with the above theories of gravity, one needs to derive the expressions for various cosmological observables within their framework. Apart from this, it has also been shown that the Riemannian structure of the space-time in GR may disappear when it is averaged over large scales as done in cosmology \cite{zala1,zala2}. Therefore, it is of interest to check what effects do the non-Riemannian variables have on observables such as cosmological distances.
	
	In cosmology, light rays incoming from distant sources are used for distance measurements. That is, we trace the optical ray bundles (congruence of null geodesics) coming from a given source back in time (from us towards the source) to determine the distance to this source. The cross sectional area, $ A $, of this light bundle is proportional to the angular diameter distance, $ d_A $ \cite{sasaki,peebles,ellis1,ellis2},
	\begin{equation}
		A \propto d^2_A
	\end{equation} 
	
	Since the expansion scalar is related to the cross sectional `area' of the null congruences, we can derive an equation for the angular diameter distance in terms of the geometrical quantities. To derive a formula for the angular diameter distance in metric-affine geometry, we restrict ourselves to four dimensions $  (n = 4)  $ so that the $ (n-2) $-volume is simply the cross sectional area, $ A $, of the congruence. The question of determining the path of physical particles has been a subject of debate in the literature (see \cite{obu2,hehl2,adam,heyde,pue1,pue2,del1,del2,del3} and the references therein). A na\"ive choice would be taking nr-autoparallel curves as the photon paths. However, it has been argued that if the physical particles do not possess the micro-structure (e.g. spin) that will couple with the non-Riemannian variables of the metric-affine geometry, then one should choose their paths to be simple r-autoparallels \cite{obu2}. To compare these choices, we will derive the distance equation for  both the cases: first assuming that photons travel on nr-autoparallel curves and then assuming that they travel on r-autoparallels. It is important to note that the true oaths of physical particles should be derived from the dynamical equations of a particular theory.
	
	\subsection{NR-Autoparallels as Photon Paths}
	For nr-autoparallels, equation \eqref{fracareachange2} becomes,
	\begin{equation}\label{fracareachange3}
		\frac{1}{A}\bar{\Df}_k A = \frac{2}{d_A}\bar{\Df}_k d_A = \bar{\Theta} - (2-m^2)N_{\alpha\beta\sigma}X^\alpha k^\beta k^\sigma - 2T_\alpha k^\alpha
	\end{equation}
	Taking a directional derivative of the above equation, we get,
	\begin{multline}\label{angdis1}
		\frac{2}{d_A}\bar{\Df}_k^2 d_A = \bar{\Df}_k \bar{\Theta} + \hf \bar{\Theta}^2 + k^\rho M_\rho N_{\alpha\beta\sigma}X^\alpha k^\beta k^\sigma - (2-m^2)\bar{\Df}_k\left( N_{\alpha\beta\sigma}X^\alpha k^\beta k^\sigma \right) \\ - 2\bar{\Df}_k \left(T_\alpha k^\alpha\right) + 2T_\alpha T_\beta k^\alpha k^\beta + \frac{(2-m^2)}{2}^2 N_{\alpha\beta\sigma}X^\alpha k^\beta k^\sigma N_{\rho\epsilon\lambda}X^\rho k^\epsilon k^\lambda  \\+ 2(2-m^2)N_{\alpha\beta\sigma}X^\alpha k^\beta k^\sigma T_\rho k^\rho - 2\bar{\Theta}T_\alpha k^\alpha - (2-m^2)\bar{\Theta}N_{\alpha\beta\sigma}X^\alpha k^\beta k^\sigma
	\end{multline}
	Using equation \eqref{einkinnull1}, this becomes,
	\begin{multline}\label{angdis2}
		\frac{2}{d_A}\bar{\Df}_k^2 d_A = \bar{\Omega}^2 - \bar{\Sigma}^2 - \bar{R}_{\alpha\beta}k^\alpha k^\beta + k^\rho M_\rho N_{\alpha\beta\sigma}X^\alpha k^\beta k^\sigma  - \bar{\nabla}_\alpha\left({N^\alpha}_{\beta\sigma}k^\beta k^\sigma\right) \\ - (3-m^2)\bar{\Df}_k\left( N_{\alpha\beta\sigma}X^\alpha k^\beta k^\sigma \right) - 2\bar{\Df}_k \left(T_\alpha k^\alpha\right) - 2 {N^\alpha}_{\beta\sigma}k^\beta k^\sigma \bar{\nabla}_\alpha k^\rho X_\rho \\+ \frac{(1-m^2)(3-m^2)}{2} N_{\alpha\beta\sigma}X^\alpha k^\beta k^\sigma N_{\rho\epsilon\lambda}X^\rho k^\epsilon k^\lambda + 2T_\alpha T_\beta k^\alpha k^\beta \\+ 2(2-m^2)N_{\alpha\beta\sigma}X^\alpha k^\beta k^\sigma T_\rho k^\rho  - 2\bar{\Theta}T_\alpha k^\alpha - (2-m^2)\bar{\Theta}N_{\alpha\beta\sigma}X^\alpha k^\beta k^\sigma 
	\end{multline}
	The above equation is valid for arbitrary space-times. In general, the distances have contributions from both the torsion and non-metricity. 
	
	The analysis of distances until now has been general and the resulting equations are quite complex. However, these equations can be simplified by considering vectorial forms of torsion and non-metricity and further assuming fixed length vectors ($ m^2 = 1 $). Then, using equations \eqref{torvec} and \eqref{nonmetvec} in equation \eqref{fracareachange3}, we get,
	\begin{equation}\label{angdisvec1}
		\frac{2}{d_A}\bar{\Df}_k d_A = \bar{\Theta} - \frac{8}{3}T_\alpha k^\alpha - \hf Q_\alpha k^\alpha
	\end{equation}
	Solving this, we get,
	\begin{equation}\label{angdisvec2}
		d_A (t) = \e^{\hf \int \left(\bar{\Theta} - \frac{8}{3}T_\alpha k^\alpha - \hf Q_\alpha k^\alpha\right)\df t}
	\end{equation}
	From the above expression, we see that the effect of the non-Riemannian variables on distances will vanish if we take an ansatz, $ Q_\alpha = -\frac{16}{3}T_\alpha $. Another way to look at this is that if one considers the cases of pure torsion ($ \boldsymbol{Q} = 0 $) and pure non-metricity ($ \boldsymbol{T} = 0 $), then the above ansatz will make the effect of pure torsion on distances phenomenologically indistinguishable from the effect of pure non-metricity. Remarkably, a very similar ansatz has been used to show a similar duality between the effects of the non-Riemannian variables previously in the literature \cite{ios,berth,capo3,sot,olm,jarv}. 
	
	\subsection{R-Autoparallels as Photon Paths}
	For r-autoparallels, the equation for the distances simplifies greatly, and we have,
	\begin{equation}\label{fracareachange4}
		\bar{\Theta} - 2T_\alpha k^\alpha = \frac{1}{A}\bar{\Df}_k A = \frac{2}{d_A}\bar{\Df}_k d_A
	\end{equation}
	Taking a directional derivative again, we get,
	\begin{equation}\label{angdis3}
		\frac{2}{d_A}\bar{\Df}^2_k d_A = \Df_k \bar{\Theta} + \hf \bar{\Theta}^2 - 2\bar{\Df}_k\left(T_\alpha k^\alpha\right) + 2T_\alpha T_\beta k^\alpha k^\beta - 2\bar{\Theta}T_\alpha k^\alpha
	\end{equation}
	Using equation \eqref{einkinnull1}, we get,
	\begin{equation}\label{angdis4}
		\frac{2}{d_A}\bar{\Df}_k^2 d_A = \bar{\Omega}^2 - \bar{\Sigma}^2 - \bar{R}_{\alpha\beta}k^\alpha k^\beta - 2\bar{\nabla}_\alpha\left(k^\alpha T_\beta k^\beta\right) + 2T_\alpha T_\beta k^\alpha k^\beta 
	\end{equation}
	where we have used the fact that for r-autoparallels, $ \bar{\Theta} = \bar{\nabla}_\alpha k^\alpha $. It is clear that the effects from non-metricity vanish.
	
	The above equations are valid for arbitrary space-times including the FLRW space-time. Solving (either of) the above equation in particular space-times will give an exact form for the angular diameter distance. An important caveat to note here is that the variable `$ t $' is simply a parameter, not the proper time. In geometries with non-metricity, these two do not match in general. However, in geometries with only torsion, we can take the parameter, $ t $, to be the proper time. We see that for nr-autoparallels, there exists a duality between the effects of non-metricity and torsion. However, in case of r-autoparallels, non-metricity effects vanish identically and the apparent duality that existed in the former case does not exist in this case. Therefore, we can conclude that the choice of nr-autoparallels as the paths of physical particles plays a role in this apparent duality. In any case, the above analysis tells us that the the possibility of using cosmological distance data to detect any signatures of non-metricity in the universe is limited, no matter the choice for photon paths.
	
	\section{Summary and Discussion} \label{sec-discuss}
	In this paper, we presented the kinematics of a congruence of timelike and null curves in geometries with torsion and non-metricity. To do so, we first derived an equation for the evolution of the separation vector between two infinitesimally close curves in the congruence. Using this equation, we found the so called evolution tensor that governs the fractional rate of change of the deviation vector. In GR, this evolution tensor is simply the gradient of the tangent vector ($ n $-velocity of the curves). However, in non-Riemannian geometry, we found that this tensor has contributions from both non-metricity and torsion. We also derived the deviation equation and found that both the non-Riemannian variables contribute to a relative acceleration between initially parallel curves.
	
	Decomposing the evolution tensor into its irreducible parts, we derived the expressions for the three kinematic variables, namely, expansion, rotation and shear. We separated the non-Riemannian and Riemannian terms in these expressions to see the modifications caused by the non-Riemannian nature of the geometry explicitly. Using these expressions for the kinematic variables, we derived their evolution equations. These included the most general forms of the Raychaudhuri equation and the Sachs optical equation\footnote{To the best of our knowledge, the equation for shear of timelike curves, and all the three kinematic equations for null curves in a metric-affine geometry are appearing in the literature for the first time. The equations for expansion and rotation of timelike curves for geometries with torsion and non-metricity have been derived previously in \cite{ios,iosth}. However, the form in which they appear in this paper is still novel since our. One must also be cautious of the differences between the definitions of the kinematic variables here and in \cite{ios,iosth}. The kinematic variables here have been defined using the full evolution tensor, $ B_{\alpha\beta} $, instead of simply the velocity gradient.}. 
	
	Using the expressions for the expansion scalar, we showed that it does not retain its Riemannian interpretation of being equal to the fractional rate of change of the cross sectional volume (or area) of the congruence. Rather, the expansion is related to the (rate of change of) cross section via a scalar made purely of non-Riemannian terms. This scalar can be made to vanish by assuming an ansatz relationship between the two non-Riemannian variables without necessarily putting them to zero. Further, analysing the so called Frobenius theorem in the context of metric-affine geometries, we showed that if a congruence is assumed to be hypersurface orthogonal, the associated rotation tensor can be determined completely in terms of the distortion tensor. That is, the rotation of hypersurface orthogonal congruences is a purely non-Riemannian effect.
	
	As a quick application of the generalised Sachs optical equation, we derived a general formula for angular diameter distances in arbitrary space-times. Assuming the photon trajectories to be nr-autoparallel curves and taking vectorial forms of both the torsion and non-metricity, we showed that the effect of these two variables on the distances can be made to vanish by taking a simple ansatz on them. Another way of looking at this is that the cases of pure torsion and pure non-metricity are phenomenologically indistinguishable given this ansatz holds. This so called duality between the effects of these variables has previously been reported in the literature \cite{ios,berth,capo3,sot,olm,jarv,klemm}. However, this duality disappears if one assumes the photon trajectories to be r-autoparallels instead of nr-autoparallels. Therefore, we attribute the duality that we found to the choice of nr-autoparallels as photon trajectories.
	
	To increase the applicability of the results in this paper further, we presented specific forms of the Raychaudhuri equation and the Sachs optical equation relevant to various non-Riemannian theories of gravity. The Raychaudhuri equation has extensive applications such as in the study of gravitational collapse, singularity theorems, relativistic cosmology, black hole mechanics, space-time thermodynamics, etc. On the other hand, Sachs optical equation is used in the derivation of cosmological distances and in weak lensing theory. It has also been employed in the study of gravitational radiation. It is our belief that the results of this paper will enable one to study the aforementioned topics in the frameworks of these extensions of GR. Apart from these classical issues, since metric-affine gravity takes into account the gravitational effects of the microstructure (quantum properties) of physical particles (e.g., spin, hypercharge), the results in this paper might prove useful in studies of quantum gravity which has indeed been one of the motivations behind metric-affine theories of gravity \cite{hehl1}. 
	
	\subsection*{Acknowledgements}
	We thank Sai Madhav Modumudi for numerous helpful discussions and comments on the manuscript. We also thank Damianos Iosifidis for useful comments.

	\bibliographystyle{unsrtnat}	
	\bibliography{reinftq}	

\begin{thebibliography}{68}
\providecommand{\natexlab}[1]{#1}
\providecommand{\url}[1]{\texttt{#1}}
\expandafter\ifx\csname urlstyle\endcsname\relax
  \providecommand{\doi}[1]{doi: #1}\else
  \providecommand{\doi}{doi: \begingroup \urlstyle{rm}\Url}\fi

\bibitem[Einstein(1915)]{ein1}
Albert Einstein.
\newblock {The Field Equations of Gravitation}.
\newblock \emph{Sitzungsber. Preuss. Akad. Wiss. Berlin (Math. Phys. )},
  1915:\penalty0 844--847, 1915.

\bibitem[Einstein(1916)]{ein2}
Albert Einstein.
\newblock {The Foundation of the General Theory of Relativity}.
\newblock \emph{Annalen Phys.}, 49\penalty0 (7):\penalty0 769--822, 1916.
\newblock \doi{10.1002/andp.19163540702}.

\bibitem[Will(2018)]{will}
Clifford~M. Will.
\newblock \emph{Theory and Experiment in Gravitational Physics}.
\newblock Cambridge University Press, 2 edition, 2018.
\newblock \doi{10.1017/9781316338612}.

\bibitem[Schouten(1954)]{schout}
J.A. Schouten.
\newblock \emph{Ricci-Calculus: An Introduction to Tensor Analysis and Its
  Geometrical Applications}.
\newblock Die Grundlehren der mathematischen Wissenschaften in
  Einzeldarstellungen. Springer, 1954.
\newblock ISBN 9783642056925.
\newblock URL \url{https://books.google.com/books?id=ExMZAQAAIAAJ}.

\bibitem[Weyl(1918)]{weyl1}
H.~Weyl.
\newblock {Gravitation and electricity}.
\newblock \emph{Sitzungsber. Preuss. Akad. Wiss. Berlin (Math. Phys. )},
  1918:\penalty0 465, 1918.

\bibitem[Cartan(1922{\natexlab{a}})]{cart0}
E.~Cartan.
\newblock Sur les équations de la gravitation d'einstein.
\newblock \emph{Journal de Mathématiques Pures et Appliquées}, 1:\penalty0
  141--204, 1922{\natexlab{a}}.
\newblock URL \url{http://eudml.org/doc/235383}.

\bibitem[Cartan(1922{\natexlab{b}})]{cart1}
{\'E}lie Cartan.
\newblock Sur les {\'e}quations de structure des espaces
  g{\'e}n{\'e}ralis{\'e}s et l’expression analytique du tenseur d’einstein.
\newblock \emph{Comptes Rendus Acad{\'e}mie des Sciences}, 174:\penalty0
  1104ff, 1922{\natexlab{b}}.

\bibitem[Cartan(1923)]{cart2}
Elie Cartan.
\newblock Sur les vari\'et\'es \`a connexion affine et la th\'eorie de la
  relativit\'e g\'en\'eralis\'ee (premi\`ere partie).
\newblock \emph{Annales scientifiques de l'\'Ecole Normale Sup\'erieure}, 3e
  s{\'e}rie, 40:\penalty0 325--412, 1923.
\newblock \doi{10.24033/asens.751}.
\newblock URL \url{http://www.numdam.org/articles/10.24033/asens.751/}.

\bibitem[Cartan(1924)]{cart3}
Elie Cartan.
\newblock Sur les vari\'et\'es \`a connexion affine, et la th\'eorie de la
  relativit\'e g\'en\'eralis\'ee (premi\`ere partie) {(Suite)}.
\newblock \emph{Annales scientifiques de l'\'Ecole Normale Sup\'erieure}, 3e
  s{\'e}rie, 41:\penalty0 1--25, 1924.
\newblock \doi{10.24033/asens.753}.
\newblock URL \url{http://www.numdam.org/articles/10.24033/asens.753/}.

\bibitem[Cartan(1925)]{cart4}
Elie Cartan.
\newblock Sur les vari\'et\'es \`a connexion affine, et la th\'eorie de la
  relativit\'e g\'en\'eralis\'ee (deuxi\`eme partie).
\newblock \emph{Annales scientifiques de l'\'Ecole Normale Sup\'erieure}, 3e
  s{\'e}rie, 42:\penalty0 17--88, 1925.
\newblock \doi{10.24033/asens.761}.
\newblock URL \url{http://www.numdam.org/articles/10.24033/asens.761/}.

\bibitem[Beltr\'an~Jim\'enez et~al.(2019)Beltr\'an~Jim\'enez, Heisenberg, and
  Koivisto]{jim}
Jose Beltr\'an~Jim\'enez, Lavinia Heisenberg, and Tomi~S. Koivisto.
\newblock {The Geometrical Trinity of Gravity}.
\newblock \emph{Universe}, 5\penalty0 (7):\penalty0 173, 2019.
\newblock \doi{10.3390/universe5070173}.

\bibitem[Capozziello et~al.(2022)Capozziello, De~Falco, and Ferrara]{capo4}
Salvatore Capozziello, Vittorio De~Falco, and Carmen Ferrara.
\newblock {Comparing equivalent gravities: common features and differences}.
\newblock \emph{Eur. Phys. J. C}, 82\penalty0 (10):\penalty0 865, 2022.
\newblock \doi{10.1140/epjc/s10052-022-10823-x}.

\bibitem[Hayashi and Shirafuji(1979)]{haya}
Kenji Hayashi and Takeshi Shirafuji.
\newblock {New General Relativity}.
\newblock \emph{Phys. Rev. D}, 19:\penalty0 3524--3553, 1979.
\newblock \doi{10.1103/PhysRevD.19.3524}.
\newblock [Addendum: Phys.Rev.D 24, 3312--3314 (1982)].

\bibitem[Maluf(2013)]{maluf}
J.~W. Maluf.
\newblock {The teleparallel equivalent of general relativity}.
\newblock \emph{Annalen Phys.}, 525:\penalty0 339--357, 2013.
\newblock \doi{10.1002/andp.201200272}.

\bibitem[Aldrovandi and Pereira(2013)]{aldro}
Ruben Aldrovandi and Jos\'e~Geraldo Pereira.
\newblock \emph{{Teleparallel Gravity}: {An Introduction}}.
\newblock Springer, 2013.
\newblock ISBN 978-94-007-5142-2, 978-94-007-5143-9.
\newblock \doi{10.1007/978-94-007-5143-9}.

\bibitem[Bahamonde et~al.(2023)Bahamonde, Dialektopoulos, Escamilla-Rivera,
  Farrugia, Gakis, Hendry, Hohmann, Levi~Said, Mifsud, and Di~Valentino]{baha}
Sebastian Bahamonde, Konstantinos~F. Dialektopoulos, Celia Escamilla-Rivera,
  Gabriel Farrugia, Viktor Gakis, Martin Hendry, Manuel Hohmann, Jackson
  Levi~Said, Jurgen Mifsud, and Eleonora Di~Valentino.
\newblock {Teleparallel gravity: from theory to cosmology}.
\newblock \emph{Rept. Prog. Phys.}, 86\penalty0 (2):\penalty0 026901, 2023.
\newblock \doi{10.1088/1361-6633/ac9cef}.

\bibitem[Ferraris and Kijowski(1982)]{ferr}
M.~Ferraris and J.~Kijowski.
\newblock {ON THE EQUIVALENCE OF THE RELATIVISTIC THEORIES OF GRAVITATION}.
\newblock \emph{Gen. Rel. Grav.}, 14:\penalty0 165--180, 1982.
\newblock \doi{10.1007/BF00756921}.

\bibitem[Nester and Yo(1999)]{nest}
James~M. Nester and Hwei-Jang Yo.
\newblock {Symmetric teleparallel general relativity}.
\newblock \emph{Chin. J. Phys.}, 37:\penalty0 113, 1999.

\bibitem[Capozziello et~al.(2021)Capozziello, Finch, Said, and Magro]{capo1}
Salvatore Capozziello, Andrew Finch, Jackson~Levi Said, and Alessio Magro.
\newblock {The 3+1 formalism in teleparallel and symmetric teleparallel
  gravity}.
\newblock \emph{Eur. Phys. J. C}, 81\penalty0 (12):\penalty0 1141, 2021.
\newblock \doi{10.1140/epjc/s10052-021-09944-6}.

\bibitem[Obukhov and Puetzfeld(2014)]{obu1}
Yuri~N. Obukhov and Dirk Puetzfeld.
\newblock Conservation laws in gravity: A unified framework.
\newblock \emph{Phys. Rev. D}, 90:\penalty0 024004, Jul 2014.
\newblock \doi{10.1103/PhysRevD.90.024004}.
\newblock URL \url{https://link.aps.org/doi/10.1103/PhysRevD.90.024004}.

\bibitem[Hehl et~al.(1995)Hehl, McCrea, Mielke, and Ne'eman]{hehl1}
Friedrich~W. Hehl, J.~Dermott McCrea, Eckehard~W. Mielke, and Yuval Ne'eman.
\newblock {Metric affine gauge theory of gravity: Field equations, Noether
  identities, world spinors, and breaking of dilation invariance}.
\newblock \emph{Phys. Rept.}, 258:\penalty0 1--171, 1995.
\newblock \doi{10.1016/0370-1573(94)00111-F}.

\bibitem[Ellis et~al.(2012)Ellis, Maartens, and MacCallum]{ellisbook}
George F.~R. Ellis, Roy Maartens, and Malcolm A.~H. MacCallum.
\newblock \emph{Relativistic Cosmology}.
\newblock Cambridge University Press, 2012.
\newblock \doi{10.1017/CBO9781139014403}.

\bibitem[Poisson(2004)]{poisson}
Eric Poisson.
\newblock \emph{A Relativist's Toolkit: The Mathematics of Black-Hole
  Mechanics}.
\newblock Cambridge University Press, 2004.
\newblock \doi{10.1017/CBO9780511606601}.

\bibitem[{Schneider} et~al.(1992){Schneider}, {Ehlers}, and
  {Falco}]{schneidbook}
Peter {Schneider}, J{\"u}rgen {Ehlers}, and Emilio~E. {Falco}.
\newblock \emph{{Gravitational Lenses}}.
\newblock Springer, 1992.
\newblock \doi{10.1007/978-3-662-03758-4}.

\bibitem[Jacobson(1995)]{jacob}
Ted Jacobson.
\newblock {Thermodynamics of space-time: The Einstein equation of state}.
\newblock \emph{Phys. Rev. Lett.}, 75:\penalty0 1260--1263, 1995.
\newblock \doi{10.1103/PhysRevLett.75.1260}.

\bibitem[Hawking and Ellis(1973)]{hawkell}
S.~W. Hawking and G.~F.~R. Ellis.
\newblock \emph{The Large Scale Structure of Space-Time}.
\newblock Cambridge Monographs on Mathematical Physics. Cambridge University
  Press, 1973.
\newblock \doi{10.1017/CBO9780511524646}.

\bibitem[{Kar} and {Sengupta}(2007)]{kar}
Sayan {Kar} and Soumitra {Sengupta}.
\newblock {The Raychaudhuri equations: A brief review}.
\newblock \emph{Pramana}, 69\penalty0 (1):\penalty0 49, July 2007.
\newblock \doi{10.1007/s12043-007-0110-9}.

\bibitem[Luz and Vitagliano(2017)]{luz}
Paulo Luz and Vincenzo Vitagliano.
\newblock Raychaudhuri equation in spacetimes with torsion.
\newblock \emph{Phys. Rev. D}, 96:\penalty0 024021, Jul 2017.
\newblock \doi{10.1103/PhysRevD.96.024021}.
\newblock URL \url{https://link.aps.org/doi/10.1103/PhysRevD.96.024021}.

\bibitem[Hensh and Liberati(2021)]{hensh}
Sudipta Hensh and Stefano Liberati.
\newblock Raychaudhuri equations and gravitational collapse in einstein-cartan
  theory.
\newblock \emph{Phys. Rev. D}, 104:\penalty0 084073, Oct 2021.
\newblock \doi{10.1103/PhysRevD.104.084073}.
\newblock URL \url{https://link.aps.org/doi/10.1103/PhysRevD.104.084073}.

\bibitem[Dey and Majhi(2022{\natexlab{a}})]{dey1}
Sumit Dey and Bibhas~Ranjan Majhi.
\newblock Kinematics and dynamics of null hypersurfaces in the einstein-cartan
  spacetime and related thermodynamic interpretation.
\newblock \emph{Phys. Rev. D}, 105:\penalty0 064047, Mar 2022{\natexlab{a}}.
\newblock \doi{10.1103/PhysRevD.105.064047}.
\newblock URL \url{https://link.aps.org/doi/10.1103/PhysRevD.105.064047}.

\bibitem[Dey et~al.(2017)Dey, Liberati, and Pranzetti]{dey2}
Ramit Dey, Stefano Liberati, and Daniele Pranzetti.
\newblock Spacetime thermodynamics in the presence of torsion.
\newblock \emph{Phys. Rev. D}, 96:\penalty0 124032, Dec 2017.
\newblock \doi{10.1103/PhysRevD.96.124032}.
\newblock URL \url{https://link.aps.org/doi/10.1103/PhysRevD.96.124032}.

\bibitem[Dey and Majhi(2022{\natexlab{b}})]{dey3}
Sumit Dey and Bibhas~Ranjan Majhi.
\newblock {Possible fluid interpretation and tidal force equation on a generic
  null hypersurface in Einstein-Cartan theory}.
\newblock \emph{Phys. Rev. D}, 106\penalty0 (10):\penalty0 104005,
  2022{\natexlab{b}}.
\newblock \doi{10.1103/PhysRevD.106.104005}.

\bibitem[{Capozziello} et~al.(2001){Capozziello}, {Lambiase}, and
  {StornaioloI}]{capo2}
S.~{Capozziello}, G.~{Lambiase}, and C.~{StornaioloI}.
\newblock {Geometric classification of the torsion tensor of space‑time}.
\newblock \emph{Annalen der Physik}, 513\penalty0 (8):\penalty0 713--727,
  August 2001.
\newblock
  \doi{10.1002/andp.2001513080310.1002/1521-3889(200108)10:8<713::AID-ANDP713>3.0.CO;2-2}.

\bibitem[{Wanas} and {Bakry}(2009)]{wanas}
M.~I. {Wanas} and M.~A. {Bakry}.
\newblock {Effect of Spin-Torsion Interaction on Raychaudhuri Equation}.
\newblock \emph{International Journal of Modern Physics A}, 24\penalty0
  (27):\penalty0 5025--5032, January 2009.
\newblock \doi{10.1142/S0217751X09046291}.

\bibitem[{Cai} et~al.(2016){Cai}, {Capozziello}, {De Laurentis}, and
  {Saridakis}]{cai}
Yi-Fu {Cai}, Salvatore {Capozziello}, Mariafelicia {De Laurentis}, and
  Emmanuel~N. {Saridakis}.
\newblock {f(T) teleparallel gravity and cosmology}.
\newblock \emph{Reports on Progress in Physics}, 79\penalty0 (10):\penalty0
  106901, October 2016.
\newblock \doi{10.1088/0034-4885/79/10/106901}.

\bibitem[Pasmatsiou et~al.(2017)Pasmatsiou, Tsagas, and Barrow]{pasma}
Klaountia Pasmatsiou, Christos~G. Tsagas, and John~D. Barrow.
\newblock Kinematics of einstein-cartan universes.
\newblock \emph{Phys. Rev. D}, 95:\penalty0 104007, May 2017.
\newblock \doi{10.1103/PhysRevD.95.104007}.
\newblock URL \url{https://link.aps.org/doi/10.1103/PhysRevD.95.104007}.

\bibitem[Speziale(2018)]{spez}
Simone Speziale.
\newblock Raychaudhuri and optical equations for null geodesic congruences with
  torsion.
\newblock \emph{Phys. Rev. D}, 98:\penalty0 084029, Oct 2018.
\newblock \doi{10.1103/PhysRevD.98.084029}.
\newblock URL \url{https://link.aps.org/doi/10.1103/PhysRevD.98.084029}.

\bibitem[Lobo et~al.(2015)Lobo, Barreto, and Romero]{lobo}
I.~P. Lobo, A.~B. Barreto, and C.~Romero.
\newblock {Space-time singularities in Weyl manifolds}.
\newblock \emph{Eur. Phys. J. C}, 75\penalty0 (9):\penalty0 448, 2015.
\newblock \doi{10.1140/epjc/s10052-015-3671-7}.

\bibitem[Iosifidis(2019)]{iosth}
Damianos Iosifidis.
\newblock Metric-affine gravity and cosmology/aspects of torsion and
  non-metricity in gravity theories, 2019.
\newblock URL \url{https://arxiv.org/abs/1902.09643}.

\bibitem[Iosifidis et~al.(2018)Iosifidis, Tsagas, and Petkou]{ios}
Damianos Iosifidis, Christos~G. Tsagas, and Anastasios~C. Petkou.
\newblock Raychaudhuri equation in spacetimes with torsion and nonmetricity.
\newblock \emph{Phys. Rev. D}, 98:\penalty0 104037, Nov 2018.
\newblock \doi{10.1103/PhysRevD.98.104037}.
\newblock URL \url{https://link.aps.org/doi/10.1103/PhysRevD.98.104037}.

\bibitem[Yang et~al.(2021)Yang, Shahidi, Harko, and Liang]{yang}
Jin-Zhao Yang, Shahab Shahidi, Tiberiu Harko, and Shi-Dong Liang.
\newblock {Geodesic deviation, Raychaudhuri equation, Newtonian limit, and
  tidal forces in Weyl-type $f(Q,T)$ gravity}.
\newblock \emph{Eur. Phys. J. C}, 81\penalty0 (2):\penalty0 111, 2021.
\newblock \doi{10.1140/epjc/s10052-021-08910-6}.

\bibitem[Misner et~al.(1973)Misner, Thorne, and Wheeler]{mtw}
Charles~W. Misner, K.S. Thorne, and J.A. Wheeler.
\newblock \emph{{Gravitation}}.
\newblock W. H. Freeman, San Francisco, 1973.
\newblock ISBN 978-0-7167-0344-0, 978-0-691-17779-3.

\bibitem[Clifton et~al.(2012)Clifton, Ferreira, Padilla, and Skordis]{clifrev}
Timothy Clifton, Pedro~G. Ferreira, Antonio Padilla, and Constantinos Skordis.
\newblock {Modified Gravity and Cosmology}.
\newblock \emph{Phys. Rept.}, 513:\penalty0 1--189, 2012.
\newblock \doi{10.1016/j.physrep.2012.01.001}.

\bibitem[Nojiri et~al.(2017)Nojiri, Odintsov, and Oikonomou]{nojrev}
S.~Nojiri, S.~D. Odintsov, and V.~K. Oikonomou.
\newblock {Modified Gravity Theories on a Nutshell: Inflation, Bounce and
  Late-time Evolution}.
\newblock \emph{Phys. Rept.}, 692:\penalty0 1--104, 2017.
\newblock \doi{10.1016/j.physrep.2017.06.001}.

\bibitem[Raychaudhuri(1955)]{ray1}
Amalkumar Raychaudhuri.
\newblock {Relativistic cosmology. 1.}
\newblock \emph{Phys. Rev.}, 98:\penalty0 1123--1126, 1955.
\newblock \doi{10.1103/PhysRev.98.1123}.

\bibitem[Sachs and Bondi(1961)]{sachs1}
R.~Sachs and Hermann Bondi.
\newblock Gravitational waves in general relativity. vi. the outgoing radiation
  condition.
\newblock \emph{Proceedings of the Royal Society of London. Series A.
  Mathematical and Physical Sciences}, 264\penalty0 (1318):\penalty0 309--338,
  1961.
\newblock \doi{10.1098/rspa.1961.0202}.
\newblock URL
  \url{https://royalsocietypublishing.org/doi/abs/10.1098/rspa.1961.0202}.

\bibitem[Sachs(1962)]{sachs2}
R.~K. Sachs.
\newblock {Gravitational waves in general relativity. 8. Waves in
  asymptotically flat space-times}.
\newblock \emph{Proc. Roy. Soc. Lond. A}, 270:\penalty0 103--126, 1962.
\newblock \doi{10.1098/rspa.1962.0206}.

\bibitem[{Zalaletdinov}(1992)]{zala1}
Roustam~M. {Zalaletdinov}.
\newblock {Averaging out the Einstein equations}.
\newblock \emph{General Relativity and Gravitation}, 24\penalty0 (10):\penalty0
  1015--1031, October 1992.
\newblock \doi{10.1007/BF00756944}.

\bibitem[Zalaletdinov(1993)]{zala2}
Roustam Zalaletdinov.
\newblock {Towards a theory of macroscopic gravity}.
\newblock \emph{Gen. Rel. Grav.}, 25:\penalty0 673--695, 1993.
\newblock \doi{10.1007/BF00756937}.

\bibitem[Sasaki(1993)]{sasaki}
Misao Sasaki.
\newblock {Cosmological gravitational lens equation: Its validity and
  limitation}.
\newblock \emph{Prog. Theor. Phys.}, 90:\penalty0 753--781, 1993.
\newblock \doi{10.1143/PTP.90.753}.

\bibitem[{Peebles}(1993)]{peebles}
P.~J.~E. {Peebles}.
\newblock \emph{{Principles of Physical Cosmology}}.
\newblock Princeton University Press, 1993.

\bibitem[Ellis(1971)]{ellis1}
G.~F.~R. Ellis.
\newblock {Relativistic cosmology}.
\newblock \emph{Proc. Int. Sch. Phys. Fermi}, 47:\penalty0 104--182, 1971.
\newblock \doi{10.1007/s10714-009-0760-7}.

\bibitem[{Ellis}(2009)]{ellis2}
George F.~R. {Ellis}.
\newblock {Republication of: Relativistic cosmology}.
\newblock \emph{General Relativity and Gravitation}, 41\penalty0 (3):\penalty0
  581--660, March 2009.
\newblock \doi{10.1007/s10714-009-0760-7}.

\bibitem[Obukhov and Puetzfeld(2021)]{obu2}
Yuri~N. Obukhov and Dirk Puetzfeld.
\newblock {Demystifying autoparallels in alternative gravity}.
\newblock \emph{Phys. Rev. D}, 104\penalty0 (4):\penalty0 044031, 2021.
\newblock \doi{10.1103/PhysRevD.104.044031}.

\bibitem[Hehl(1971)]{hehl2}
F.W. Hehl.
\newblock How does one measure torsion of space-time?
\newblock \emph{Physics Letters A}, 36\penalty0 (3):\penalty0 225--226, 1971.
\newblock ISSN 0375-9601.
\newblock \doi{https://doi.org/10.1016/0375-9601(71)90433-6}.
\newblock URL
  \url{https://www.sciencedirect.com/science/article/pii/0375960171904336}.

\bibitem[Adamowicz(1975)]{adam}
\&~Trautman~A. Adamowicz, W.
\newblock The principle of equivalence for spin.
\newblock \emph{Bulletin de l'Academie Polonaise des Sciences Serie des
  Sciences, Mathematiques, Astronomiques et Physiques}, 23\penalty0
  (3):\penalty0 339--342, 1975.
\newblock URL
  \url{https://inis.iaea.org/search/search.aspx?orig_q=RN:10431161}.

\bibitem[Von Der~Heyde(1975)]{heyde}
P.~Von Der~Heyde.
\newblock The equivalence principle in the $ u_4 $ theory of gravitation.
\newblock \emph{Lettere al Nuovo Cimento (1971-1985)}, 14:\penalty0 250--252,
  1975.
\newblock \doi{https://doi.org/10.1007/BF02745635}.
\newblock URL \url{https://link.springer.com/article/10.1007/BF02745635}.

\bibitem[Puetzfeld and Obukhov(2013)]{pue1}
Dirk Puetzfeld and Yuri~N. Obukhov.
\newblock Equations of motion in gravity theories with nonminimal coupling: A
  loophole to detect torsion macroscopically?
\newblock \emph{Phys. Rev. D}, 88:\penalty0 064025, Sep 2013.
\newblock \doi{10.1103/PhysRevD.88.064025}.
\newblock URL \url{https://link.aps.org/doi/10.1103/PhysRevD.88.064025}.

\bibitem[Puetzfeld and Obukhov(2014)]{pue2}
Dirk Puetzfeld and Yuri~N. Obukhov.
\newblock Equations of motion in metric-affine gravity: A covariant unified
  framework.
\newblock \emph{Phys. Rev. D}, 90:\penalty0 084034, Oct 2014.
\newblock \doi{10.1103/PhysRevD.90.084034}.
\newblock URL \url{https://link.aps.org/doi/10.1103/PhysRevD.90.084034}.

\bibitem[Delhom et~al.(2020)Delhom, Lobo, Olmo, and Romero]{del1}
Adria Delhom, Iarley~P. Lobo, Gonzalo~J. Olmo, and Carlos Romero.
\newblock {Conformally invariant proper time with general non-metricity}.
\newblock \emph{Eur. Phys. J. C}, 80\penalty0 (5):\penalty0 415, 2020.
\newblock \doi{10.1140/epjc/s10052-020-7974-y}.

\bibitem[Beltr\'an~Jim\'enez and Delhom(2020)]{del2}
Jose Beltr\'an~Jim\'enez and Adri\`a Delhom.
\newblock {Instabilities in metric-affine theories of gravity with higher order
  curvature terms}.
\newblock \emph{Eur. Phys. J. C}, 80\penalty0 (6):\penalty0 585, 2020.
\newblock \doi{10.1140/epjc/s10052-020-8143-z}.

\bibitem[Delhom(2021)]{del3}
Adri\`a Delhom.
\newblock \emph{{Theoretical and Observational Aspecs in Metric-Affine Gravity:
  A field theoretic perspective}}.
\newblock PhD thesis, Valencia U., 2021.

\bibitem[Berthias and Shahid-Saless(1993)]{berth}
Jean-Paul Berthias and Bahman Shahid-Saless.
\newblock {Torsion and nonmetricity in scalar - tensor theories of gravity}.
\newblock \emph{Class. Quant. Grav.}, 10:\penalty0 1039--1044, 1993.
\newblock \doi{10.1088/0264-9381/10/5/020}.

\bibitem[Capozziello et~al.(2007)Capozziello, Cianci, Stornaiolo, and
  Vignolo]{capo3}
S.~Capozziello, R.~Cianci, C.~Stornaiolo, and S.~Vignolo.
\newblock {f(R) gravity with torsion: The Metric-affine approach}.
\newblock \emph{Class. Quant. Grav.}, 24:\penalty0 6417--6430, 2007.
\newblock \doi{10.1088/0264-9381/24/24/015}.

\bibitem[Sotiriou(2009)]{sot}
Thomas~P. Sotiriou.
\newblock {f(R) gravity, torsion and non-metricity}.
\newblock \emph{Class. Quant. Grav.}, 26:\penalty0 152001, 2009.
\newblock \doi{10.1088/0264-9381/26/15/152001}.

\bibitem[Olmo(2011)]{olm}
Gonzalo~J. Olmo.
\newblock {Palatini Approach to Modified Gravity: f(R) Theories and Beyond}.
\newblock \emph{Int. J. Mod. Phys. D}, 20:\penalty0 413--462, 2011.
\newblock \doi{10.1142/S0218271811018925}.

\bibitem[J\"arv et~al.(2018)J\"arv, R\"unkla, Saal, and Vilson]{jarv}
Laur J\"arv, Mihkel R\"unkla, Margus Saal, and Ott Vilson.
\newblock {Nonmetricity formulation of general relativity and its scalar-tensor
  extension}.
\newblock \emph{Phys. Rev. D}, 97\penalty0 (12):\penalty0 124025, 2018.
\newblock \doi{10.1103/PhysRevD.97.124025}.

\bibitem[Klemm and Ravera(2020)]{klemm}
Dietmar~Silke Klemm and Lucrezia Ravera.
\newblock {Einstein manifolds with torsion and nonmetricity}.
\newblock \emph{Phys. Rev. D}, 101\penalty0 (4):\penalty0 044011, 2020.
\newblock \doi{10.1103/PhysRevD.101.044011}.

\end{thebibliography}
	%\addcontentsline{toc}{section}{References}	
	
\end{document}